\documentclass[10pt,conference]{IEEEtran}
\IEEEoverridecommandlockouts
\usepackage{caption}

\let\oldcaption\caption
\def\caption#1{\oldcaption{\lowercase{#1}}}

\usepackage[linesnumbered,algoruled,boxed,lined]{algorithm2e}
\usepackage{algpseudocode}
\usepackage{makecell}
\usepackage{amsmath}
\usepackage[numbers]{natbib}
\usepackage{multirow}
\usepackage{booktabs}
\usepackage{hyperref}
\usepackage{diagbox}
\usepackage{amssymb}
\hypersetup{
    colorlinks=true,
    linkcolor=blue,
    filecolor=magenta,
    urlcolor=magenta,
}
\usepackage[all]{hypcap}
\usepackage{booktabs}
\usepackage{url}
\usepackage{xcolor}
\usepackage{tikz}
\usepackage{diagbox}
\usepackage{listings}
\usepackage{comment} 
\usepackage{graphicx}
\usepackage[framemethod=TikZ]{mdframed}
\usepackage{tcolorbox}
\usepackage{amsthm}

\theoremstyle{definition}

\tcbset{
    colback=gray!10, 
    colframe=gray, 
    coltext=black, 
    boxsep=5pt, 
    arc=0mm, 
    top=0mm, bottom=0mm, 
    left=1mm, right=1mm, 
    boxrule=0.5mm, 
    toptitle=1mm, bottomtitle=1mm, 
    titlerule=0mm, 
}
\definecolor{deepgreen}{rgb}{0.0, 0.4, 0.0}

\definecolor{codegreen}{rgb}{0,0.6,0}
\definecolor{codegray}{rgb}{0.5,0.5,0.5}
\definecolor{codepurple}{rgb}{0.58,0,0.82}
\definecolor{backcolour}{rgb}{0.95,0.95,0.92}

\lstdefinestyle{mystyle}{
  backgroundcolor=\color{backcolour}, commentstyle=\color{codegreen},
  keywordstyle=\color{magenta},
  numberstyle=\tiny\color{codegray},
  stringstyle=\color{codepurple},
  basicstyle=\ttfamily\footnotesize,
  breakatwhitespace=false,         
  breaklines=true,                 
  captionpos=b,                    
  keepspaces=true,                 
  numbers=left,                    
  numbersep=5pt,                  
  showspaces=false,                
  showstringspaces=false,
  showtabs=false,                  
  tabsize=2
}
\usepackage{xurl}

\definecolor{mycolor}{RGB}{194, 214, 236}

\newcounter{finding}

\newcommand{\find}[1]{
\begin{tcolorbox}[leftrule=0.5mm,toprule=0mm,bottomrule=0mm,left=0.7pt,right=0.7pt,top=0.2pt,bottom=0.2pt]
\em #1
\end{tcolorbox}
}

\newcounter{result}

\makeatletter
\g@addto@macro{\@algocf@init}{\SetKwInOut{Parameter}{Parameters}}
\makeatother

\newcommand*\circled[1]{\tikz[baseline=(char.base)]{
            \node[shape=circle, draw, inner sep=0.2pt] (char) {\textcolor{black}{#1}};}}

\newcommand{\tech}{\mbox{\textsc{MalGraph}}}   
 
\begin{document}

\begin{sloppypar}
\title{An Analysis of Malicious Packages in Open-Source Software in the Wild}

\author{
    \IEEEauthorblockN{Xiaoyan Zhou\IEEEauthorrefmark{1}, Ying Zhang\IEEEauthorrefmark{1}, Wenjia Niu\IEEEauthorrefmark{1}, 
    Jiqiang Liu\IEEEauthorrefmark{1}, Haining Wang\IEEEauthorrefmark{2}, 
    Qiang Li\IEEEauthorrefmark{1}\IEEEauthorrefmark{3}\thanks{\IEEEauthorrefmark{3}Qiang Li is the corresponding author; email: liqiang@bjtu.edu.cn}}
\IEEEauthorblockA{\IEEEauthorrefmark{1}\textit{School of Cyberspace Security, Beijing Jiaotong University, China}  \\
\IEEEauthorrefmark{2} \textit{Department of Electrical and Computer Engineering, Virginia Polytechnic Institute and State University, USA} 
}  
}
    \vspace{-10mm}

\maketitle
\thispagestyle{plain}  
\begin{abstract}
The open-source software (OSS) ecosystem suffers from security threats caused by malware.
However, OSS malware research has three limitations: a lack of high-quality datasets, a lack of malware diversity, and a lack of attack campaign contexts. 
In this paper, we first build the largest dataset of 24,356 malicious packages from online sources, then propose a knowledge graph to represent the OSS malware corpus and conduct malware analysis in the wild.
Our main findings include 
(1) it is essential to collect malicious packages from various online sources because their data overlapping degrees are small;
(2) despite the sheer volume of malicious packages, many reuse similar code, leading to a low diversity of malware;
(3) only 28 malicious packages were repeatedly hidden via dependency libraries of 1,354 malicious packages, and dependency-hidden malware has a shorter active time;
(4) security reports are the only reliable source for disclosing the malware-based context.

\end{abstract}


\section{Introduction}

Open-source software (OSS) ecosystems provide a large number of reusable packages, libraries, tools, and processes for software project development.
Meanwhile, OSS ecosystems suffer from a variety of security threats and risks, called software supply chain (SSC) attacks~\cite{ladisa2023sok}, where malicious code is injected into the development, build, and release of artifacts.
Consequently, OSS malware, software packages containing malicious code segments that execute unauthorized/malicious behaviors, plays a central role in SSC attacks.
Malicious packages initially deceive developers and users to download and install them and then execute subsequent behaviors, such as implanting backdoors~\cite{ssc_backdoor}, stealing sensitive information~\cite{ssc_credential}, and downloading and executing payloads without user permission (e.g., cryptominers~\cite{ssc_crypt}).
For example, the `Fallguys'~\cite{ssc_fallguy} is a malicious package that steals sensitive data stored on a computer, extracts the AWS token key, and sends it to a remote server. 
This software package has been downloaded more than 345 times, posing a security threat to users. 
Therefore, studying the OSS malicious package and building a defense for safeguarding the OSS ecosystem is highly needed.

OSS malware has been studied in recent years, with a focus on the detection and analysis of malicious packages~\cite{SejiaAdria2022Machinelearn, zhang2020cyber, qian2022malicious, ferreira2021containing, vu2023bad}.
However, the existing malware research has several limitations.
First, traditional malware studies~\cite{bailey2007automated, bayer2009scalable, jang2011bitshred, vx_underground, rokon2020sourcefinder} provide compiled malware in the binary format rather than OSS malicious packages. 
Most OSS malicious packages use interpreted code, and there are no binary-level malware samples.
Second, the existing OSS malware datasets~\cite{duan2020towards, guo2023empirical} are limited in size and from a single OSS ecosystem: 1,195 malicious packages in the GitHub repository~\cite{duan2020towards}, and 2,915 PyPI malicious packages in~\cite{guo2023empirical}.
Third, many malicious packages are unavailable because the registry removes the packages. 
As per our investigation, the missing rate is close to 39.27\%.
Fourth, the diversity of malicious packages is not considered in the existing datasets. 
A large number of malicious packages does not imply malware diversity.
We only obtain 153 similar code groups from the PyPI malware dataset (Section~\ref{sec:sub:diversity}).

To address those limitations, we first collect a large-scale and high-quality dataset of OSS malware packages from different online sources. 
Our observation is that security professionals detect and disclose malicious packages. 
Many commercial security enterprises have their own security blogs to report their findings and malware analysis. 
Those security enterprises regularly scan the newly added software packages in the OSS ecosystems and discover new and unknown malicious packages.
Once publicly disclosed, the analysis information about a malicious package is freely available to a third party. Based on this observation, we have collected  24,356 OSS malicious packages.
To our knowledge, this is the largest OSS malware dataset.

Further, we leverage the knowledge graph ({\tech}) to represent the OSS malware dataset, where each node is a malicious package, and each edge represents a package relationship.
We propose four types of edges (Section~\ref{sec:graph}): \textit{duplicated}, \textit{dependency}, \textit{similar}, and \textit{co-existing} relationships.
Malicious packages are grouped into a subgraph via those relationships.
{\tech} uses a subgraph to represent the OSS malware-based context (e.g., an attack campaign, or malware diversity). 
For example, a threat actor will release a malicious package multiple times to launch an attack campaign, and those packages may be grouped together.
We also use the subgraph to characterize an attack campaign toward OSS ecosystems and provide a quantitative analysis of the OSS malware-based context.

\textbf{Findings}. 
Based on our analysis results, there are four major learned lessons. 
(1) There is a lack of industry-wide and industry-academia collaboration, limiting the data quality of malicious packages. 
There is little data overlap, and many studies are conducted independently.
OSS malware analysis research is still in the infancy stage, and collecting malicious packages from various online sources is imperative. 
(2) Despite an increasing number of OSS malicious packages, their code diversity is low. 
A typical attack campaign is a repeated attempt in which attackers reuse similar code to distribute different malicious packages to the registry.
(3) A small number of malicious packages are repeatedly concealed within other malicious packages, with some being hidden over 400 times.
Yet, using dependency libraries to hide malicious packages has a short average active period (10.5 days), and the repetitive distribution based on similar code remains active for approximately 45.16 days.
(4) Malcious packages often lack context about how and who released them, but security reports reveal missing SSC attack campaigns. 
The reported attack campaign discloses thousands of suspicious IP addresses and malicious domains.


\section{OSS Malware Collection}

A reliable and high-quality dataset of OSS malware is essential for the security community.  
In this section, we first highlight the distinguishing characteristics of OSS malware and then detail the methodology for collecting the OSS malware dataset.

\subsection{OSS Malware}

\begin{figure}[!t]
 \begin{center}
    \includegraphics[width=3.4in ]{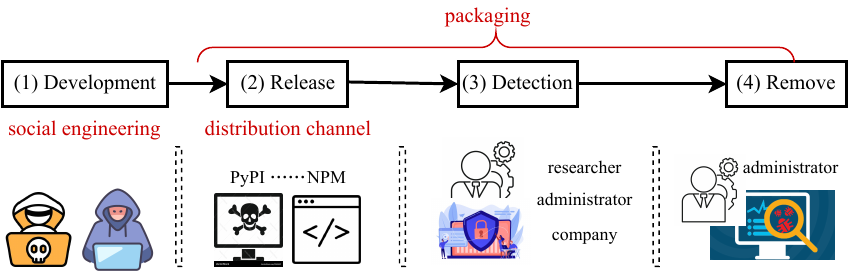}
      \caption{A life cycle of a malicious package: (1) the development, (2) releasing malware to OSS, (3) detection, and (4) removing the malware.}
     \label{fig:life:cycle}
    \end{center}
    \vspace{-5mm}
\end{figure}

OSS malware differs from conventional malware in three aspects.  
(1) \textit{Distribution Channels}: OSS and conventional malware use different channels for distribution. OSS malware primarily spreads through package registries (e.g., PyPI and NPM), while conventional malware is typically delivered through phishing emails, malicious attachments, infected software, or compromised websites.
(2) \textit{Packaging}: OSS malware is packaged as a software component within a registry, whereas conventional malware is usually delivered as compiled binaries.
This difference in packaging allows OSS malware to be integrated into software libraries, while conventional malware varies by architecture and compiler, resulting in distinct binary samples for different systems.
(3) \textit{Social Engineering}: OSS malware uses a different social engineering tactic to deceive users into downloading and installing it. 
It often uses names, versions, and descriptions similar to trusted software packages (e.g., typosquatting and combosquatting), making it difficult for victim users to differentiate between legitimate and malicious software.

In terms of the above different places, the life cycle of a malicious package can be characterized by four distinct phases: development, release, detection, and removal, as illustrated in Figure~\ref{fig:life:cycle}.
The first phase starts when an adversary or a hacker develops a malicious package.
During this phase, social engineering tactics are employed to disguise the malware and make it appear as a legitimate package.  
The second phase occurs when the adversary releases the malicious package into the OSS ecosystem. 
This stage determines the distribution channel through which the malware spreads, with the package being added to a registry and remaining in this form throughout the subsequent phases.  
In the third phase, the malicious package is detected by a security researcher, a security company, or a package management administrator.
If the malicious package is first found by a security researcher or a security company, they will notify the package manager administrator.
Finally, the package administrator removes the malicious package from the registry.
A typical malware's life can be summarized as the sequence: \{development$\rightarrow$release$\rightarrow$detection$\rightarrow$removal\}.

The distinguishing characteristics of OSS malware lead to two practical challenges in malware collection:  
\begin{itemize}  
    \item Once package managers take down malware, it becomes inaccessible from package registries such as NPM, PyPI, and RubyGems.  
    \item Ethical and business considerations prevent companies and security professionals from sharing malware datasets publicly.  
\end{itemize}

\subsection{Data Collection Methodology}
\label{sec:sub:data}

\begin{figure*}
    \centering
    \includegraphics[width=4.7 in ]{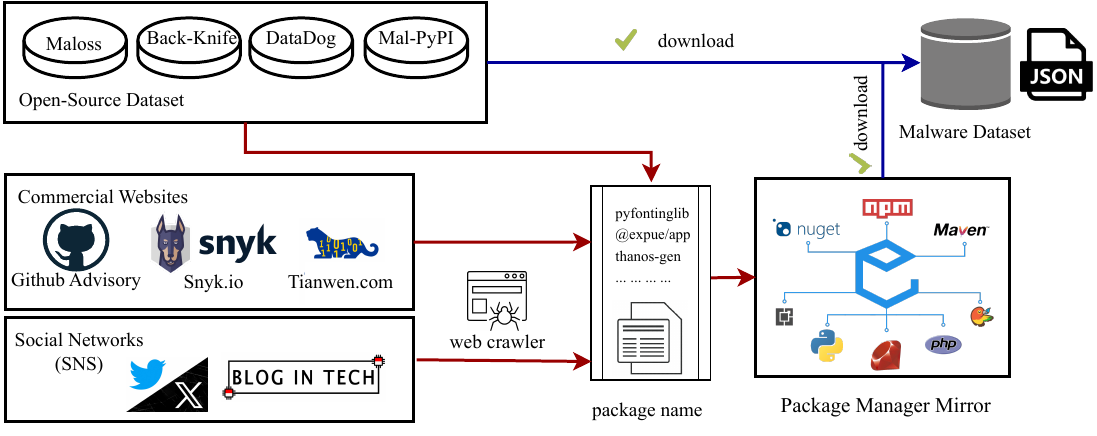}
    \caption{Data collection methodology for OSS malicious packages. (1) We center around scattered online sources: open-source datasets, commercial websites, individual blogs, and social networks. (2) If a malicious package is available, we directly download it. If a malicious package is taken down by sources or OSS registries, we record its name and version. (3) We query the OSS registry mirrors to find unavailable packages based on names and versions.}
    \label{fig:data}
\vspace{-4mm}
\end{figure*}

To address these challenges, we propose a data collection methodology, as illustrated in Figure~\ref{fig:data}.
(1) Some known/public malicious package datasets are available, and we can merge them into a large dataset. 
(2) When malicious packages are not available, we use their names/versions to find corresponding packages. 
(3) Many package mirror registries are not properly synced with the root registry. This means that malicious packages may still exist in the mirror registry, even if they were removed from the main registry.
We can find malicious packages in the mirror registry via package names/versions.

\textbf{Open-source Dataset}.
To collect malicious packages for the study, we conducted an in-depth investigation of relevant literature about the OSS ecosystem and the SCC attack. 
Typically, we have found four available open-source datasets for malicious packages, including the Maloss dataset~\cite{duan2020towards}, Backstabber-Knife dataset~\cite{backstabbers-online}, Mal-PyPI dataset~\cite{guo2023empirical}, and DataDog dataset~\cite{datadog}. 
For those open-source datasets, we directly download malicious packages if they are available.

A practical problem for those open-source datasets is that the package number is inconsistent with the number in the research work.
Mal-PyPI Dataset~\cite{guo2023empirical} claimed it collected 4,500+ PyPI packages, but we only obtained 2,915 malicious packages.
The remaining 1,500+ packages are unavailable, and we cannot obtain their package names and versions.
To deal with the practical issue, we record the package name/version from the open-source dataset if malicious packages are unavailable.
We use mirror registries to search package names/versions to obtain their packages.

\textbf{Commercial Websites and Individual Blogs}.
Security researchers or professionals often share malicious package information and publish security reports on webpages.
Typically, a webpage about a security analysis report contains several parts: (1) when or how to find the malicious package, (2) its malicious behaviors or objectives, (3) the package name and version, and (4) a possible useful indicator of compromised (IOC). 
For instance, Snyk.io reported the backdoor packages~\cite{Liran_ruby_backdoor} in the RubyGem ecosystem, where its name is `bootstrap-sass' and its version is `3.2.0.3'.
We investigated an in-depth investigation of popular websites related to OSS ecosystems and SCC attacks. 
From these websites, we manually picked those related to OSS ecosystems and added them to the list, including Snyk.io~\cite{snyk-online}, Tianwen~\cite{tianwen-online}, GitHub Advisory~\cite{github-advisory}, Phylum~\cite{phylum-online}, Socket~\cite{socket-online}, and individual blogs~\cite{in-sscblog-ivan, in-sscblog-jor, in-sscblog-cata}.

\textbf{Social Networks (SNS)} are the source for collecting malicious packages.  
From our observation, X (Twitter) is the most active platform with the highest update frequency for finding malicious packages.
However, the most significant limitation is that online sources (commercial security websites, social networks, and individual blogs) don't provide any package to the public because they treat malicious packages as important assets and do not share the dataset.

To deal with the limitation, we use package names/versions as keywords to search multiple mirror registries.
Malicious packages are removed from the package registry but are still available in package mirror registries.
The reason is that the package mirror registries are not synced with the root registry, and there is a time gap between them.
If they exist, we download them from the package registry mirrors.
If they do not exist in the registry mirrors, we only have the package name and version without its package.
Specifically, we use 5 NPM mirrors~\cite{npm_taobao, npm_cnpmjs, npm_aliyun, npm_ustc, npm_huawei}, 12 PyPI mirrors~\cite{pypi_tsinghua, pypi_aliyun, pypi_douban, pypi_ustc, pypi_tencent, pypi_huawei, pypi_bfsu, pypi_163, pypi_sustech, pypi_rstudio, pypi_unpad, pypi_kakao}, and 6 RubyGems mirrors~\cite{ruby_taobao, ruby_tsinghua, ruby_hust, ruby_aliyun, ruby_sysu, ruby_sdutlinux} to search for malicious packages.

Table~\ref{tab:source} lists sources and sizes of initial malicious packages. 
In total, we have collected 24,356 malicious packages, 13,932 of which are available and 10,424 of which are unavailable.
``Unavailable’’ indicates we only have package names/versions, while ``Available’’ indicates we have the malware package. 
The dataset covers 10 OSS ecosystems, including PyPI, NPM, RubyGems, Maven, Cocoapods, SourceForge, Docker, Composer, NuGet, and Rust.
One straightforward observation is that most malicious packages come from PyPI and NPM, popular OSS ecosystems where criminals and adversaries prefer to launch attacks. 
In addition, we find that the overlapping degree between SNS and commercial websites is high. The reason is that many SNS accounts point to the same organizations of commercial websites and individual blogs.
Note that those SNS tweets are the same as the webpage content on commercial websites and individual blogs.

\setlength{\textfloatsep}{4mm}   
 \begin{table}[!t] \small
    \centering
    \caption{Source and size of initial malicious packages.}
    \label{tab:source}
    \begin{tabular}{c c  c c  }
    \toprule
    
   Category        &   Data Source           & \makecell[c]{ Unavailable  }   &  \makecell[c]{ Available }   \\ \hline 
    \multirow{3}*{Academia}  & Backstabber-Knife~\cite{backstabbers-online}       &0   & 5,937    \\ 
                                      &  Maloss~\cite{duan2020towards}          &0   & 1,223   \\  
                                      &  Mal-PyPI~\cite{guo2023empirical}     & 0  & 2,915     \\ 
                                     \hline
     \multirow{7}*{Industry}      &  GitHub Advisory~\cite{github-advisory}      & 166  & 13      \\ 
                              &   Snyk.io~\cite{snyk-online}       & 1,160  & 380   \\     
                              &   Tianwen~\cite{tianwen-online}         & 1,769   & 1,382  \\
                             &  DataDog~\cite{datadog}    & 0    & 1,387         \\  
                             &  Phylum~\cite{phylum-online}     & 6,606  &692          \\  
                             &  Socket~\cite{socket-online}     & 664  & 0          \\  
                            
       & Blogs~\cite{in-sscblog-ivan, in-sscblog-jor, in-sscblog-cata}       & 59   & 3     \\ 
          
                            \toprule
     \multicolumn{2}{c}{Total}            & 10,424   &  13,932       \\ 
    \toprule        
    \end{tabular}
\end{table}

\section{{\tech}: Malicious Package Graph}
\label{sec:graph}

\begin{figure}
    \centering
    \includegraphics[width=3in ]{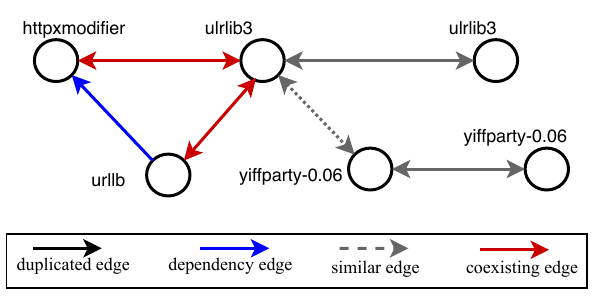}
    \caption{{\tech}: one example of OSS malicious package group.}
    \label{fig:graph}
\end{figure}

This section proposes a knowledge graph ({\tech}) to represent the malicious packages and their relationships.
Figure~\ref{fig:graph} shows a malicious package graph from our collected dataset.  
A graph is denoted as $G= (V, E)$, where the $V$ is the set of nodes, and $E$ is the set of edges. 
Each malicious package is stored as a node in the graph. 
Specifically, there are seven attributions for the node, including the ID, the package name, the package version, the source, the hash value, and the ecosystem.
We propose four types of edges to represent the relationship between two packages, including duplicated, dependency, similar, and co-existing relationships.
Specifically, we use Neo4j~\cite{neo4j_GenAI} to implement the knowledge graph.
A graph stores interlinked nodes as a knowledge graph, where we use a graph-structured data model to analyze malicious packages and their relationships.

\subsection{Duplicated Edge}

\textit{Duplicated edge} indicates two malicious packages are identical packages.
We determine whether two malicious packages belong to the same package via the following process.
If two malicious packages from different sources have identical names and versions, they are classified as duplicate packages.
Next, if malicious packages are available, their hash (e.g., MD5) is calculated to further verify the duplicate relationship.
Specifically, we leverage the Hashlib~\cite{hashlib_use} library to calculate the SHA256 of the malware code.
For instance, `acookie-1.0.0' from Maloss~\cite{maloss} and `acookie-1.0.0' from Backstabber-Knife~\cite{backstabbers-online} have the same name, version and hash value, denoted as the duplicate relationship $e($`acookie-1.0.0', `acookie-1.0.0'$)$.
In short, a duplicated edge characterizes the same malicious package from different sources, indicating the overlap of malware collection.

\begin{figure}[!t]
    \centering
    \includegraphics[width=3.2in ]{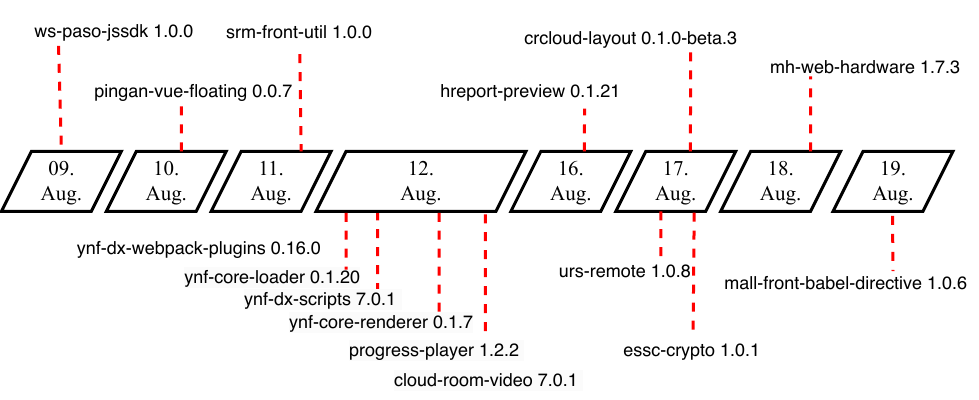}
    \caption{Example of a repeating attack~\cite{Phylum_continue_npm}: In August 2023, several subsequent malware packages were released in NPM.
    On August 9, the first malware package was released; on August 12, the ongoing campaign contained 6 different similar malware packages; between August 17 and 19, the attackers released 5 malware packages.    }
    
    \label{fig:back:2}
\end{figure}

\subsection{Similar Edge} 
\label{ch:Similar_Edge}

\textit{Similar Edge} indicates two malicious packages that share a similar source code base but have different signatures.  
The rationale behind the similar edge is that malware distributors reuse the same malware code to execute a series of repetitive attempts targeting OSS ecosystems. 
However, once a prior malware package is taken down, the same name cannot be reused, leading to different signatures.   
As a result, malware distributors must use a different package name for the subsequent attack. Over time, the same distributor may release multiple malicious packages, often based on an identical code base. 
For example, as shown in Figure~\ref{fig:back:2}, a total of 15 malware packages were released between August 9 and 19.
Note that we cannot determine whether similar packages are the result of copycat behavior or originate from the same organization or author.  
Particularly, a similar edge represents a repeated attempt derived from the same malware code base.

We determine a similar relationship between two malicious packages via the following process. 
(1) Unpacking a malware package into a folder and finding all source code files (.js, .py, .rb).
(2) Sorting all files in the same order (file names) and merging source code into a whole file. 
(3) Splitting the whole file into multiple code snippets, where each snippet has a fixed token length. 
The embedding model (512 tokes in CodeBERT-base~\cite{feng2020codebert1}) converts each snippet into a vector.
We concatenate all vectors into a large vector to represent the whole package code. 
(4) The K-Means algorithm clusters malware packages into similar groups based on the large vectors.
The minimum similarity threshold for the K-means algorithm is set to 0.7, and each subgraph must contain at least two points. 
After clustering, the silhouette score is calculated using the Scikit-learn library~\cite{scikit}. 
If the silhouette score of a cluster is less than 0.3, it is removed, indicating poor clustering performance. 
(5) We build a similar edge between two malicious packages in the same group.
The average intra-group similarity of all subgraphs in the dataset reaches 99.9\%.

\subsection{Dependency Edge}

\begin{figure}[!t]
 \begin{center}
    \includegraphics[width=2.4in ]{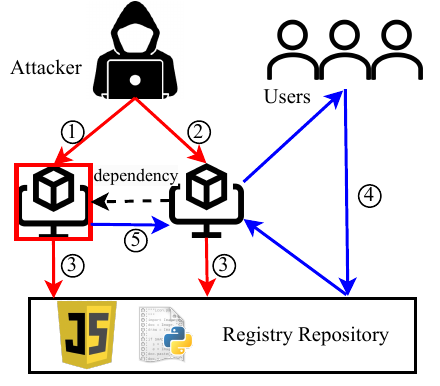}
    \caption{A dependent-hidden attack: the red line is the attack method based on the dependency relationship, and the blue line is the download process of users.}
    \label{fig:dependency}
    \end{center}
\vspace{-4mm}
\end{figure}

\textit{Dependency Edge} indicates a malicious package relies on another malicious package.
Malware distributors release multiple malicious packages, some of which, pretending to be innocent, use a dependency malicious library to hide their malicious behavior.
The benefit is that the malicious package can be shielded by an innocent package.
Figure~\ref{fig:dependency} plots the dependent-hidden attack. 
\circled{1} 
The malware developers first create a package with malicious codes to the registry repository and then create a second one. 
\circled{2} The second package, depending on the first one, seems like a legitimate package without malicious behaviors.
\circled{3} The malware distributor distributes two packages to OSS registries. 
\circled{4} The users may download and install the second package from the OSS registry. 
\circled{5} In the installation stage, the first package is automatically downloaded to the project, and the attack is triggered.
For instance, `loglib-modules' and `pygrata-utils' are two malicious packages that rely on a malicious package called `pygrata'~\cite{Ax_python}, denoted as the dependency relationship $e($`loglib-modules', `pygrata'$)$ and $e($`pygrata-utils', `pygrata'$)$.

We determine a dependency relationship between two malicious packages via the following process. 
(1) We first extract all package names and versions in the malware dataset.
(2) For the metadata, we use dependency libraries stored in the configuration file, e.g., `*.json' for NPM and `requirement.txt' for PyPI.
Note that malicious packages may claim legitimate packages as their dependency libraries.
If a dependency library is matched, we put them as the same dependent-hidden attack.
(3) For the source code, we use all malicious package names to find the matched position in the source file. 
If matched, we extract the code segment within a 100-character window.
We use regular expressions (Table~\ref{tab:regex_patterns}) to determine whether the code segment contains a dependency library from other malicious packages.
(4) We provide manual inspection over those dependency libraries to remove false positives, e.g., code comments.
Note that our matching approach belongs to the exact string-matching algorithm.

\begin{table}[!t] \small
\centering
\caption{Regular expressions for dependent-hidden libraries in source code.}
\begin{tabular}{l}
\toprule
\multicolumn{1}{c}{\textbf{Pattern}} \\ \hline
\textit{import [\textbackslash{}w./]*? from ['\textbackslash{}"][\textbackslash{}w.-]+['\textbackslash{}"]} \\ 
\textit{from ['\textbackslash{}"][\textbackslash{}w.-]+['\textbackslash{}"] import .*?}\\ 
\textit{import [\textbackslash{}w./]* from [\textbackslash{}w.-]+} $\boldsymbol{\mid}$$\boldsymbol{\mid}$ \textit{from [\textbackslash{}w.-]+ import .*?}  \\ 
\textit{import ['\textbackslash{}"][\textbackslash{}w.-]+['\textbackslash{}"]}   $\boldsymbol{\mid}$$\boldsymbol{\mid}$ \textit{import ['\textbackslash{}"][\textbackslash{}w.-]+['\textbackslash{}"]}\\ 
\textit{import [\textbackslash{}w.-]+} $\boldsymbol{\mid}$$\boldsymbol{\mid}$ 
\textit{import \textbackslash{}./[\textbackslash{}w.-]+} $\boldsymbol{\mid}$$\boldsymbol{\mid}$ 
\textit{import \textbackslash{}../[\textbackslash{}w.-]+} \\ 
\textit{import [\textbackslash{}w.]*\textbackslash{}b[\textbackslash{}w.-]+[\textbackslash{}b]} $\boldsymbol{\mid}$$\boldsymbol{\mid}$ 
\textit{from [\textbackslash{}w.-]+[\textbackslash{}w.]+ import .*?}  \\ 
\textit{(const$|$let$|$var) .* = require(['\textbackslash{}"][\textbackslash{}w.-]+['\textbackslash{}"])}  \\
\textit{require(['\textbackslash{}"][\textbackslash{}w.-]+['\textbackslash{}"])} $\boldsymbol{\mid}$$\boldsymbol{\mid}$ \textit{https?://[\textbackslash{}\textasciicircum{}\textbackslash{}s]+}\\ 
\textit{require(["\textbackslash{}'][\textasciicircum"\textbackslash{}']+(?:/[\textasciicircum"\textbackslash{}']+)*/["\textbackslash{}'])}\\

\toprule
\end{tabular}

\label{tab:regex_patterns}
\end{table}

\subsection{Co-existing Edge}

\textit{Co-existing Edge} indicates that two malicious packages co-exist in the same security report. 
Malicious packages cannot describe the context around how the malware is used and by whom. 
Hence, the security report is the evidence that reveals the attack campaign behind the malicious package.
For example, the security report~\cite{Bill_malicious} claimed that an adversary author `Lolip0p' developed malicious packages called `Colorslib',  `httpslib' and `libhttps', where the co-existing relationship $e($`Colorslib', `httpslib'$)$, $e($`Colorslib', `libhttps'$)$, and $e($`httpslib', `libhttps'$)$.
Another example is the security report~\cite{Phylum_5943_pypi}, where 5,943 malicious packages were released as a register flood attack campaign towards the PyPI registry.
The attacker released 5,943 malicious packages in a short time.

We determine a co-existing relationship between two malicious packages if they co-exist in the same security report.
A dataset with the security report is essential for building the co-existing edge in the malicious package graph.
However, there is no public dataset with the security report.

To address this practical challenge, we propose a web crawler to collect the security analysis reports from the Internet. The web crawler is implemented by the Scrapy framework~\cite{scrapy}.
(1) We use the commercial websites and individual blogs (Section~\ref{sec:sub:data}) as the initial seed for finding the security analysis reports.
(2) We leverage the search engine (Google) to find similar URLs to the initial URL.
(3) We use the web crawler to collect the security analysis reports from the Internet.
(4) We manually filter out irrelevant web pages and keep the security reports about malicious packages.
Through the manual approach, we have found 1,366 reports that provide security analysis about malicious packages. 
Table~\ref{tab:report:source} lists details of the online websites and security reports.
Our dataset contains the largest number of security reports about OSS malicious packages.

\begin{table}[!t] 
    \centering
    \caption{Source of security analysis reports. }
    \label{tab:report:source}
    \begin{tabular}{c c   c   }
    \toprule
    Category   &   Website   \#          & Report  \#            \\ \hline
    Technical Community    &      16       &  516      \\
    Commercial org.  &      15             &  545    \\
    News        &  4        &  143                   \\
    Individual   &      3             & 95                     \\
    Official    &     1              & 24            \\
    Other   &       29           & 43             \\  \hline
    Total & 68 & 1,366                   \\
    \toprule
    \end{tabular}
\end{table}


\section{Dataset and {\tech} Validity}

There is no established ground truth for OSS malware or their relationships.  
As a result, it is impossible to provide a systematic evaluation of the data collection approach (Section~\ref{sec:sub:data}) or {\tech} (Section~\ref{sec:graph}). 
Instead, we conducted controlled experiments and performed manual inspections to guarantee the reproducibility of our results.

\subsection{Malware Dataset Validity}

\textbf{Malware Correctness}. 
(1) \textit{Manual Inspection.}
We assume that online sources are reliable and have a good reputation in the security community.
If a legitimate package is falsely reported as a malicious package, it may be added to contain our dataset. 
Due to the large number of malicious packages, manually inspecting each package via the reserved engineering technique is impractical.
We use a simple rule to remove the false positives: if the root register does not remove a package, it is not a malicious package.
(2) \textit{Controlled experiment.} 
We conducted five sampling experiments from our collected dataset.  
In each experiment, we randomly selected 100 malicious packages. 
We utilized Semgrep~\cite{semgrep} and GuardDog~\cite{guarddog}, two commercial state-of-the-art tools, to scan the sampled packages. 
Although these tools may produce a high number of false positives, we manually inspected the scanning results to verify whether the packages were indeed malicious. 
The experiment demonstrated that 100\% sampling packages belong to OSS malware.

\textbf{Malware Dataset Transparency}.
To guarantee the reproducibility of analysis results, we provide the malware dataset to the community.
Due to page limitations, we present only a summary of datasets in the paper, while the detailed dataset is available in the GitHub repository.
Note that malicious packages are not available in the GitHub repository because of ethical considerations, i.e., script kiddies. 
In contrast, we share all malicious packages (Table~\ref{tab:source}) via a private GitHub repo, and applicants need to send an email request to join this repo. We believe that a high-quality and complete OSS malware dataset is an important source for our community. With it, many researchers can avoid repeating such challenging and tedious work in the SSC field.

\subsection{{\tech} Validity}

\textbf{{\tech} Correctness}.
There are several concerns about the validity of the {\tech}.
(1) \textit{False Positives}.
There are two causes bringing about false positives.
First, the similarity calculation may be inaccurate if two packages use similar codes but belong to two different groups.
There is no ground truth dataset to validate the similarity relationship.  
Second, the reports may contain malicious packages not belonging to the same attack campaign.
In this case, we provide a manual inspection to filter out the false positives.
Given a cluster or a report, we manually inspect its content to determine whether it is a false positive. 
(2) \textit{False Negatives}.
There are two causes for false negatives: unavailable packages and missing security reports. 
So far, our malware dataset (Table~\ref{tab:source}) contains the largest number of OSS malicious packages.
Our report dataset (Table~\ref{tab:report:source}) contains the largest number of security reports. 
In future work, we will continue to find and collect new malicious packages and security reports to improve the {\tech} coverage.

\textbf{{\tech} Transparency}.
To guarantee the reproducibility of analysis results, we provide the security report dataset and the knowledge graph to the community.
We list all package groups (manual labeling) so researchers can identify which package to use in the dataset. 
Our analysis contains all malicious packages from Table~\ref{tab:source}.


\begin{table*}[!t] \small
    \centering
    \caption{The overlapping matrix of all sources.  \textbf{B.K} represents Backstabber-Knife~\cite{backstabbers-online}; \textbf{M.} represents Maloss~\cite{duan2020towards}; \textbf{M.D} represents Mal-PyPI Dataset~\cite{guo2023empirical}; \textbf{G.A} represents GitHub Advisory~\cite{github-advisory}; \textbf{S.i} represents Snyk.io~\cite{snyk-online}; \textbf{T.} represents Tianwen~\cite{tianwen-online}; \textbf{D.D} represents DataDog~\cite{datadog};  \textbf{P.} represents Phylum~\cite{phylum-online}; \textbf{So.} represents Socket~\cite{socket-online}; \textbf{I.B} represents Individual Blog~\cite{in-sscblog-jor}.}
    \begin{tabular}{l|ccc || ccccccc}
    \hline
         & \makecell[c]{\textbf{B.K}\\(5,937)} & \makecell[c]{\textbf{M.}\\(1,223)} & \makecell[c]{\textbf{M.D} \\(2,915)}
         &\makecell[c]{\textbf{G.A}\\(179)} & \makecell[c]{\textbf{S.i}\\(1,540)} & \makecell[c]{\textbf{T.}\\(3,151)}& \makecell[c]{\textbf{D.D}\\(1,387)} & \makecell[c]{\textbf{P.}\\(7,299)} & \makecell[c]{\textbf{So.}\\(664)} & \makecell[c]{\textbf{I.B}\\(62)} \\
    \hline
    \textbf{B.K}(5,937) &   & \colorbox{mycolor}{368} & \colorbox{mycolor}{2,897}       & 0 & 3 & 36 & 7 & \colorbox{mycolor}{966} & 1 & 36\\
    \textbf{M.}(1,223) & \colorbox{mycolor}{368} &       & \colorbox{mycolor}{201}       & 6 & 1 & 69 & 0 & 0 & 3 & 6 \\
    \textbf{M.D}(2,915) & \colorbox{mycolor}{2,897}  & \colorbox{mycolor}{201} &         & 0 & 0 & 32 & 7 & \colorbox{mycolor}{918} & 0 & 0  \\\hline \hline
    \textbf{G.A}(179) & 0 & 6    & 0          &   & 0 & 0 & 0 & 0 & 0 & 1  \\
    \textbf{S.i}(1,540) & 3  & 1 & 0         & 0 &   & \colorbox{mycolor}{106} & 0 & 17 & 0 & 0  \\
    \textbf{T.}(3,151) & 36  & 69  & 32          & 0 & \colorbox{mycolor}{106} &    & 0 & \colorbox{mycolor}{272} & 2 & 0  \\
    \textbf{D.D}(1,387) & 7 & 0    & 7        & 0 & 0 & 0   &   & 15 & 0 & 1  \\
    \textbf{P.}(7,299) & \colorbox{mycolor}{966} & 0  & \colorbox{mycolor}{918}        & 0 & 17 & \colorbox{mycolor}{272} & 15 &   & 0 & 0 \\
    \textbf{So.}(664) & 1 & 3    & 0            & 0  & 0   & 2    & 0   & 0 &   & 0 \\
    \textbf{I.B}(62) & 36 & 6   & 0             & 1    & 0  & 0    & 1   & 0 & 0 &   \\

    \hline
    \end{tabular}
    \label{tab:matrix_all}
\end{table*}

\section{Insights of Malicious Package Group}
\label{sec:group}

In this section, we conduct a systematic analysis of the malicious packages. 
Specifically, we leverage {\tech} to answer the following research questions:

\noindent
\textbf{$\bullet$ RQ1 (Quality)}: What impact does ad-hoc research have on malicious package datasets?

\noindent
\textbf{$\bullet$ RQ2 (Diversity)}: How is the diversity of the OSS malicious package dataset?

\noindent
\textbf{$\bullet$ RQ3 (Dependent-hidden)}: Would dependent-hidden malware bring a great attack advantage?

\noindent
\textbf{$\bullet$ RQ4 (Malware Context)}: What is the context around the OSS malware packages?

\subsection{Dataset Quality (RQ1)}

We measure the dataset quality using the overlapping degree and the missing rate ($MR$).

\textbf{Overlapping}.
The research about the malicious package is done in an ad-hoc manner, indicating that cooperative relationships seem rare. 
We use the duplicated edge to represent the overlapping degree of different sources: if two nodes have a duplicated edge $e(u, v)$, we put them ($u, v$) into the same subgraph.
Table~\ref{tab:matrix_all} shows the overlapping matrix of all sources.
The left-upper part in the matrix represents the overlapping number between different sources from the academia.  
The right-lower part represents the overlapping number of different sources from the industry.
The left-lower/right-upper part represents the overlapping number between academia and industry.

\begin{figure}[!t]
 \begin{center}
    \includegraphics[width=0.4\textwidth]{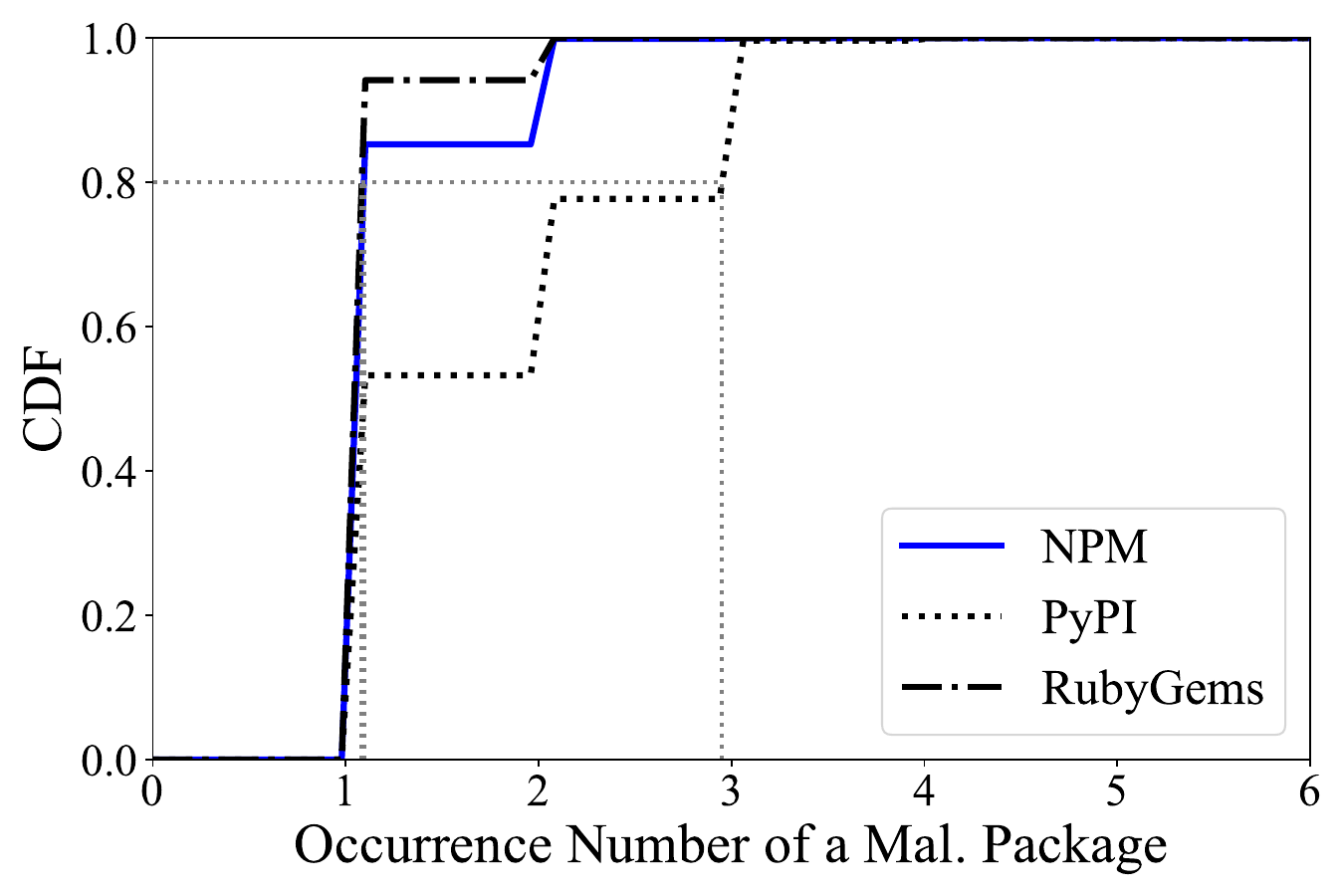}
    \end{center}
    \caption{The CDF of the number of occurrences of a malicious package among 3 OSS ecosystems: NPM, PyPI, and RubyGems.}
    \label{fig:subgraph:size}
\end{figure}

We observe that academia has a large overlapping degree (high values in left-upper/left-lower/right-upper parts), indicating a large redundancy.
When malicious packages come from academia, they are more likely to be overlapped with other sources from academia or industry. 
Obviously, Backstabber-Knife~\cite{backstabbers-online} (\textbf{B.K}) has the largest overlapping degree with other sources (respectively, 368 with \textbf{M.}, 2,897 with \textbf{M.D}, 3 with \textbf{S.i}, 36 with \textbf{T.}, 7 with \textbf{D.D}, 966 with \textbf{P.}, 1 with \textbf{So.}, 36 with \textbf{I.B}).
It is because Backstabber-Knife~\cite{backstabbers-online} is the source that integrates other datasets into its dataset.
In contrast, the industry has a small overlapping degree (small values in the right-lower part), indicating a small redundancy.
The largest overlapping degree is 272 between the Tianwen~\cite{tianwen-online} and Phylum~\cite{phylum-online}, the second largest degree is 106 between the Snyk.io~\cite{snyk-online} and Tianwen~\cite{tianwen-online}. 
Other commercial sources have a small overlapping degree, including GitHub Advisory~\cite{github-advisory}, DataDog~\cite{datadog}, and Socket~\cite{socket-online}.
The plausible reasons are as follows: (1) academia does not detect malicious packages by itself but reuses the detection result from the industry; (2) in the industry, all security corporations claim they are the first to detect malicious packages by themselves, leading to small overlapping degree.

Figure~\ref{fig:subgraph:size} shows the CDF of the number of occurrences of a malicious package in all online sources.
If a package is reported by only one source, its occurrence count is equal to 1; otherwise, it is greater than or equal to 2.
We observed that the number of occurrences for all packages is less than or equal to 4.  
In NPM and RubyGems ecosystems, 80-90\% of malicious packages appear one time and almost no package appears more than 3 times. 
Approximately 50\% of the packages on PyPI appear only once. Of the remaining, 25\% occur twice, another 25\% occur three times, while only 56 packages are observed to appear four times.

In the following analysis (Missing Rate, RQ2, RQ3, and RQ4), the dataset we used has removed the overlap.

\begin{table}[!t] \small
    \centering
    \caption{The missing rate of all sources.}
    \begin{tabular}{l c c c}
        \toprule
         Source          & \makecell{Missing Num.\# \\(Total Num.\#) } & \makecell{Local\\ $MR$ }  & \makecell{Global \\$MR$ } \\
        \midrule
        Backstabber-Knife~\cite{backstabbers-online} & 0 (5,937) & 0.00\% & 0.00\% \\
        Maloss~\cite{duan2020towards} & 0 (1,223) & 0.00\% & 0.00\% \\
        Mal-PyPI Dataset~\cite{guo2023empirical} & 0 (2,915) & 0.00\% & 0.00\%\\ \hline
        GitHub Advisory~\cite{github-advisory} & 166 (179) & 92.74\% &89.39\%\\
        Snyk.io~\cite{snyk-online} & 1,160 (1,540) & 75.32\% &75.26\% \\
        Tianwen~\cite{tianwen-online} & 1,769 (3,151) & 56.14\% &53.70\%\\
        DataDog~\cite{datadog} & 0 (1,387) & 0.00\% &0.00\%\\
        Phylum~\cite{phylum-online} & 6,606 (7,298) & 90.52\% & 80.49\%\\
        Socket~\cite{socket-online} & 664 (664) & 100\% & 99.40\%\\
        Blogs~\cite{in-sscblog-ivan, in-sscblog-jor, in-sscblog-cata} & 59 (62) & 95.16\% & 30.65\%\\
        \bottomrule
        Total & 10,424 (24,356) & &     39.27\%\\
        \bottomrule
    \end{tabular}
    \label{tab:missing}
\end{table}

\textbf{Missing Rate}.
The $MR$ reflects the dataset's quality in all sources because many malicious packages are unavailable in our dataset. 
To measure the dataset quality, we propose two metrics: local and global missing rate ($MR$).
Local $MR$ is equal to $N_{i}^m/N_i$, where $N_{i}^m$ is the number of unavailable packages in the \textit{ith} source, and $N_i$ is the number of packages in the \textit{ith} source. 
Global $MR$ is calculated as follows,
\begin{equation}
   \frac{(\Sigma_{k=1}^{N_{i}^m} x_{k})}{N{i}} \bigg| x_k = \begin{cases}
                                     1, &pkg_k \in \emptyset\\
                                    0, & pkg_k \in Set_j \quad \& \quad i \neq j
                                   \end{cases}
\end{equation}
where $pkg_k$ is missing in the \textit{ith} source, and $Set_j$ is the set of packages from other sources.
The $x_k$ is an indicator symbol: if other sources can supplement the missing package, it is assigned to 0; otherwise, it is assigned to 1.
In other words, local $MR$ reflects unavailable packages that rely on a single source, and global $MR$ reflects package unavailable from all sources.

\begin{figure}[!t]
 \begin{center}
    \includegraphics[width=3in ]{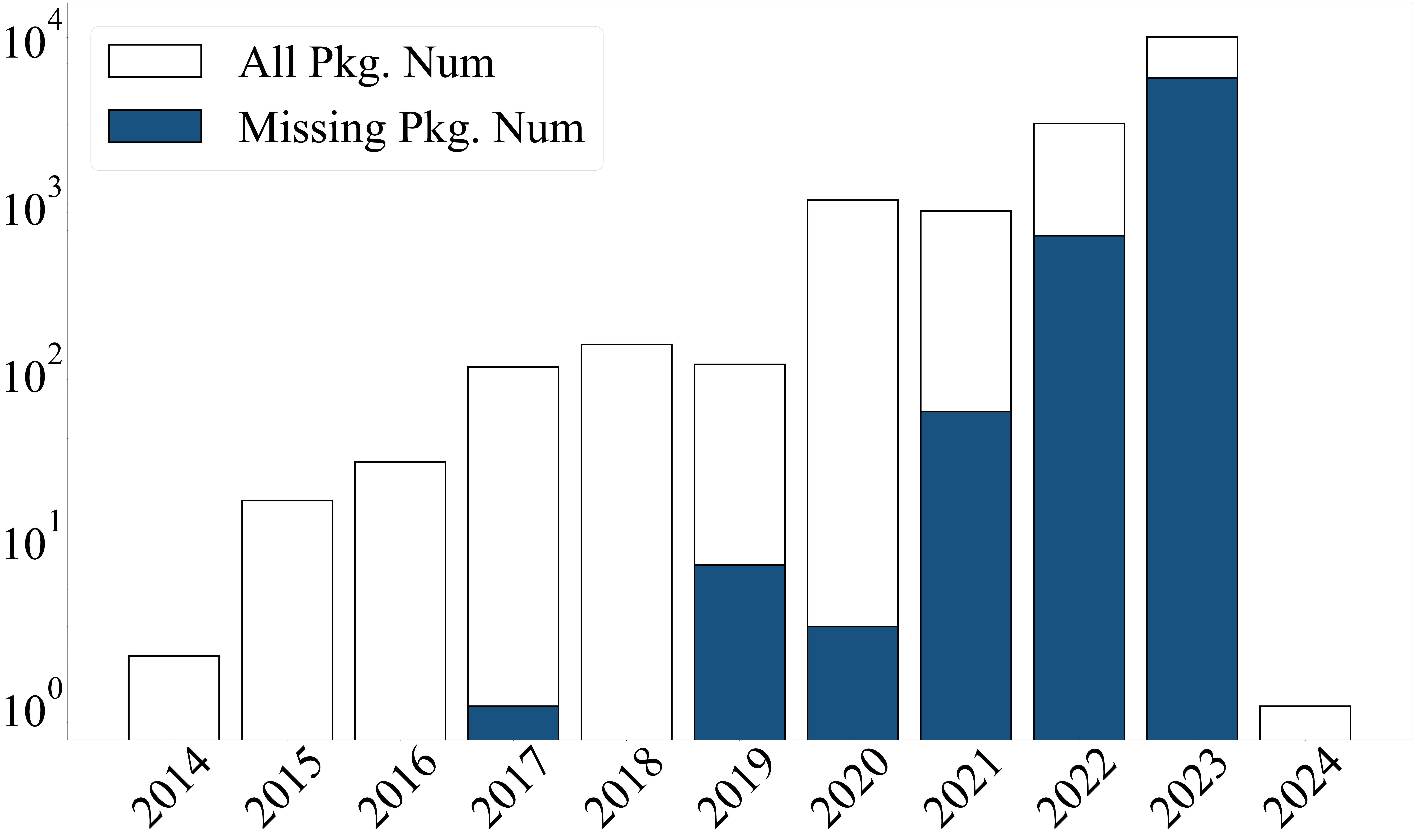}
    \caption{Package Release Timeline: all malware packages and unavailable packages.}
    \label{fig:untrack:timeline}
    \end{center}
\vspace{-4mm}
\end{figure}

Table~\ref{tab:missing} lists missing rates of all sources.
Local $MR$ in the open-source dataset (e.g., DataDog, Maloss, and Mal-PyPI) is low because we can download malicious packages directly from those sources. 
Backstabber-Knife is an important source to supplement the missing packages because it integrates malicious packages from different sources.
Global $MR$ in `Blogs' source is significantly lower than its local $MR$ because the open-source datasets supplement those missing packages. 
We observe that many sources from the industry have a large $MR$ (e.g., Socket, Phylum, Snyk.io, and Tianwen) because they do not provide malicious packages to the public.
Meanwhile, those sources have similar Global and Local $MR$, indicating that an unavailable malicious package from those sources cannot be supplemented by a different source.

\begin{figure}[!t]
    \begin{center}
       \includegraphics[width=3.2in]{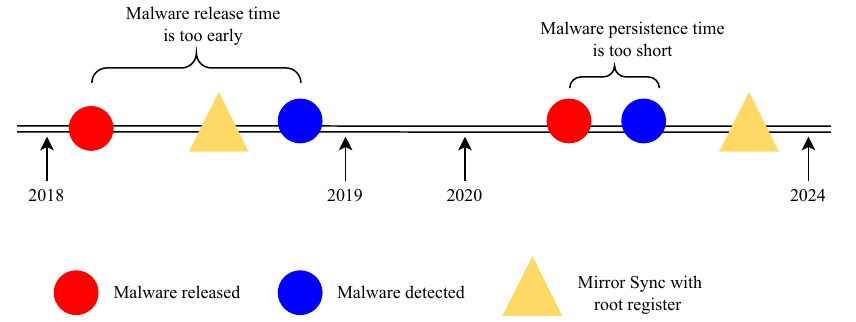}
         \caption{Two plausible causes that we cannot obtain malicious packages: (1) release time is too early, and (2) persistent period is too short.}
        \label{fig:untrack:case}
       \end{center}
\end{figure}

We further observe the release time of unavailable malicious packages. 
We query the release time to the official registry via the specific package name and version. 
Specifically, there are 24,356 malicious packages and 10,424 missing packages in our dataset.
Figure~\ref{fig:untrack:timeline} shows the release time of packages from 2014 to 2024, with the X-axis showing the release time and the Y-axis showing the package number.
The blue bar represents unavailable packages, the white bar indicates all packages.
The time of missing packages covers a span of 6 years.
We find that the number of the missing package has a peak in February 2023.
It is the reason that 5,235 malicious packages acted as the registration flooding attack against the PyPI ecosystem.
There are two reasons malicious packages were unavailable in the registry mirrors, as shown in Figure~\ref{fig:untrack:case}.
(1) The malicious package has an early release time. 
When released too early, the mirrors completed synchronization with the root registry, and the package was removed.
Figure~\ref{fig:untrack:timeline} shows the release time of missing malware packages across all sources, many of which were released 3-4 years ago.
(2) The package's persistent time is too short. 
Once a malware author releases a malicious package to the register, the administrator detects and removes it quickly. 
The sync gap between the root registry and the mirrors is larger than the persistent time of the malicious package.

\find{
\textbf{Finding 1:}
Integrating and crawling malicious packages from different online sources is necessary because of a low overlapping degree and a high missing rate.
}

\begin{table}[!t] \small
    \centering
    \caption{The details of subgraphs based on similar relationships.}
    \begin{tabular}{ c c c c c }
    \toprule
    OSS & \makecell[c]{Pkg. \\ Num.\#}  & \makecell[c]{Subgraph \\ Num.\# }  & \makecell[c]{Ave. \\ Num.\#}  & \makecell[c]{ Largest \\Size}\\
    \hline
    NPM & 2,994           &  157  & 19.07   &  827 \\
    PyPI &  4,365       &  295   & 14.80    & 829 \\
    RubyGems &  83     &  37  & 2.24    &   6 \\
    \toprule
    \end{tabular}
    \label{tab:repeat}
\end{table}

\begin{figure}[!t]
    \centering
    \includegraphics[width= 2.5in]{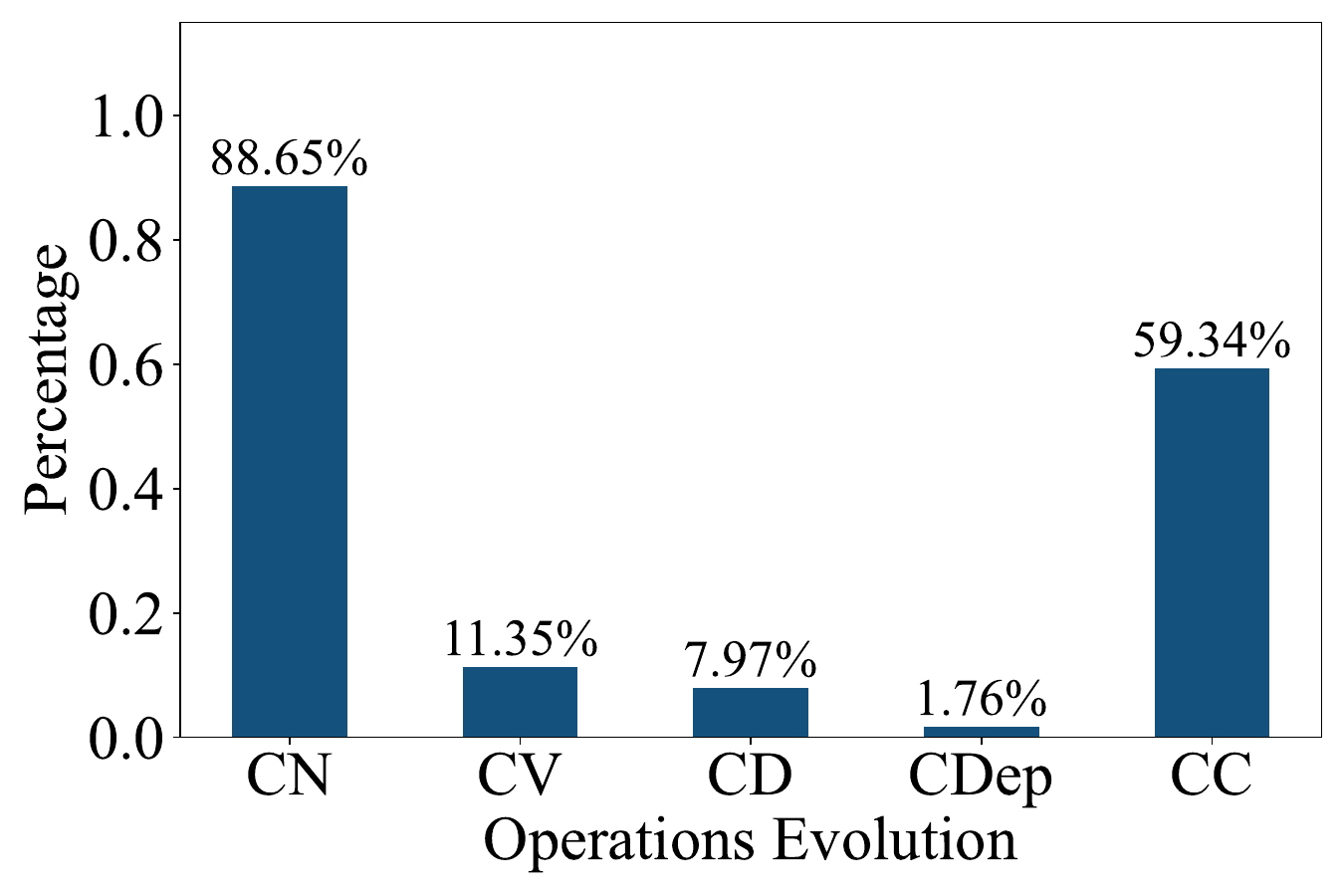}
    \caption{The operation distribution of subgraphs based on similar code.}
    \label{fig:operations evolution repeat}
    \vspace{-3mm}
\end{figure}

\subsection{Malware Diversity (RQ2)}
\label{sec:sub:diversity}

A large number of malicious packages does not imply diversity if many packages are similar. 
We use similar code bases to represent OSS malware diversity: if two nodes have a similar edge $e(u, v)$, we put them into the same subgraph.

Table~\ref{tab:repeat} lists the overall similar subgraphs based on similar code bases.
We find 157 subgraphs in NPM, 295 subgraphs in PyPI, and 37 subgraphs in RubyGems. 
The subgraph average size indicates the number of malicious packages with similar code bases.
The average number of subgraphs in PyPI contains 14.80 malicious packages; the average size in NPM is 19.07; and the subgraph in RubyGems has 2.24 packages on average.
The largest size in NPM is 827 packages, and the one in PyPI has 829 packages.
We further inspected malicious code in the largest PyPI subgraph (829). 
All packages shared the same payload that was obfuscated with variable names in Chinese.
After debugging, we found that this malware changed shortcuts to load browsers extensions. 
The browser extension replaced Crypto wallet addresses with hard-coded wallet addresses from attackers.

Despite the sheer volume of malicious packages, low software diversity brings about two implications.
First, directly using the OSS malware dataset would lead to the bias of detection models and algorithms.
If many malicious packages are centered around a few ones, the dataset is not representative, and the aggressive malicious packages dominate the dataset.
Prior works~\cite{duan2020towards, ohm2020backstabber, guo2023empirical} didn't take into account the diversity of the malicious packages.
Second,  there is a need to define malware families for OSS malware.
Traditional malware families are designed to illustrate the diversity of binary samples, which is not suitable for OSS malware.

\textbf{Social Engineering Tactics}. 
Each subgraph is a group of multiple malicious packages with a similar code base. 
When a similar malware code is reused, its life cycle can be represented as \{release $\rightarrow$ detection $\rightarrow$ removal$\rightarrow$changing$\rightarrow$ release $\dots$\}.
The removal stage is not the end of the attack; instead, the same malware is released again in the subsequent phase, creating a loop.
In each iteration, the malware must employ a social engineering tactic, as each OSS ecosystem has a security policy
for taking down malicious packages.
We use an operation to represent a social engineering tactic with multiple attack attempts.
We use $OP_i = diff(pkg_i, pkg_{i+1})$ to represent a changing operation,
where $OP_i$ is the difference between two packages, where $pkg_i$ is the prior malicious package and $pkg_{i+1}$ is the later package. 
We provide 5 kinds of operations: (1) changing name (CN); (2) changing version (CV); (3) changing description (CD); (4) changing dependency (CDep); and (5) changing source code (CC).
A changing operation is mapping to any of the five, $OP \in$ \{CN, CV, CD, CDep, CC\}.
Figure~\ref{fig:operations evolution repeat} plots the distribution of those operations among the malware groups, where the X-axis is the operation type, and the Y-axis is the percentage.
We observe that CN is the most popular operation, and CD and CDep are less frequent operations. 
The CN's percentage is 88.65\%, and the CV's percentage is 11.35\%. 
Hence, the CV operation indicates that prior malicious packages are not taken down by the registry.
In this case, attackers would use a different version to make malicious packages look more like legitimate ones.
Once a malicious package is taken down from the registry, the same package name cannot be reused. 
The CC's percentage is 59.34\%, indicating that the malware distributor will change the source code via a slight modification, e.g., use a different IP address.
We manually inspected the content of the CC operations and found that the average number of lines changed was around 0.88 lines. 
The advantage of this social engineering tactic is that it is simple and has low costs. 
The disadvantage is that today's security tools easily detect this attack campaign.
In short, the social engineering tactic for OSS malware is simple: using a different name; using a different version when the malware is not detected; and a slight change in code but keeping the original attack behavior and functions.

\begin{figure}[!t]
    \centering
    \includegraphics[width= 2.5in]{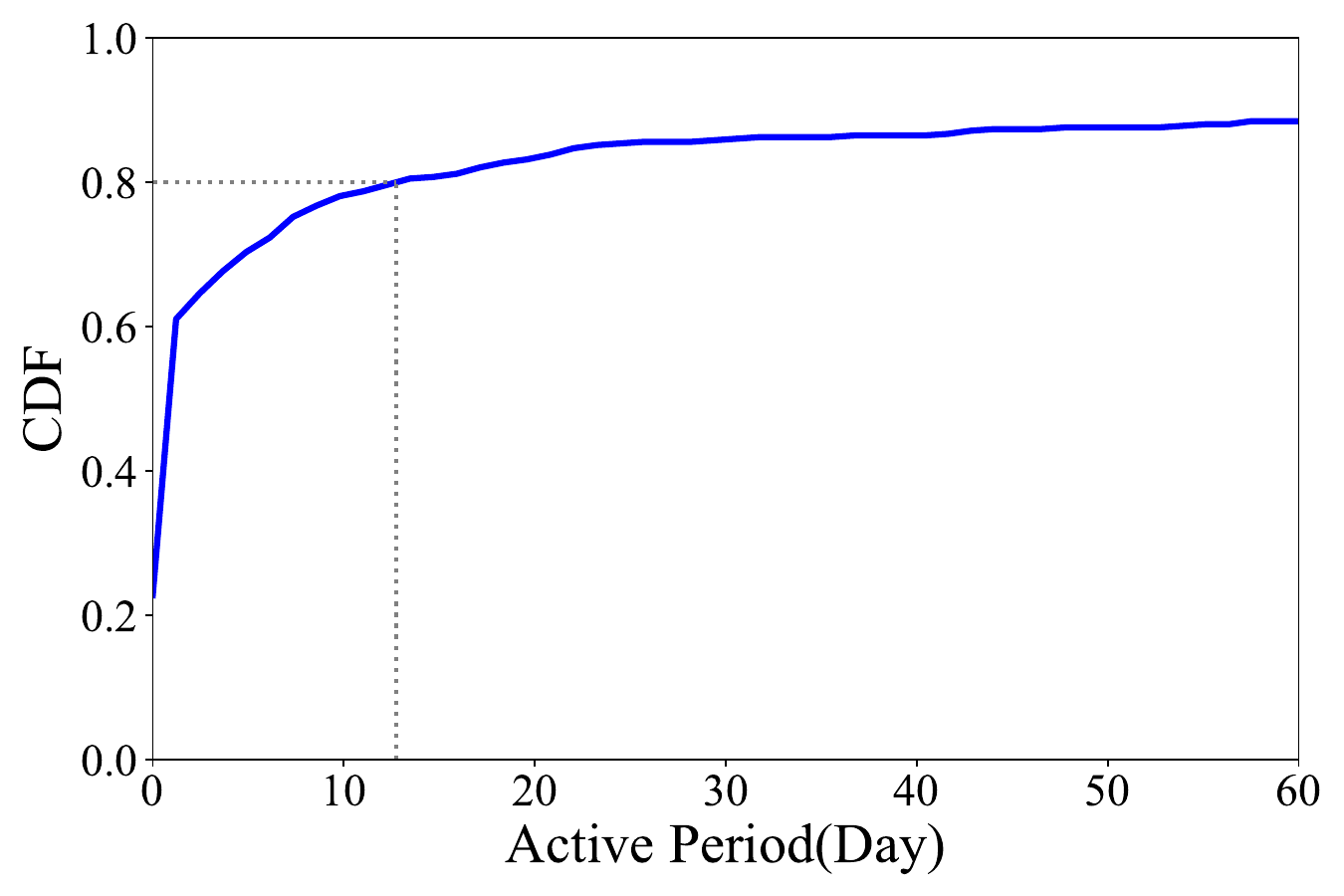}
    \caption{The active period of the subgraph based on similar code.}
    \label{fig:ap:repeat}
\end{figure}

\textbf{Active Time.}
Each subgraph has a starting time to distribute its first malware and the last time to distribute its last malware.
We use $t_{f}$ to denote the timestamp of the first malware and $t_{l}$ to denote the timestamp of the last malware.
The time span between $t_{f}$ and $t_{l}$ represents the active period of the repeated attack, denoted as $ T_{active} =  t_{l} - t_{f}$.
The active period is used to represent the time span of the subgraph, where the malware is active intermittently in the OSS ecosystem.
Figure~\ref{fig:ap:repeat} plots the CDF of the active period of the repeated attack.
It is observed that 80\% of the subgraph have an active time of less than 15 days. 
This shows that most repeated attacks are achieved by sending a large number of similar packages within a short time frame.
In addition, there are still 53 subgraphs with time intervals of more than 60 days, some of which span over 1,000 days, indicating that a similar code base has been repeatedly used over this period.

\find{
\textbf{Finding 2:}
The diversity of OSS malware is often overlooked, yet it significantly impacts security research. 
Attackers are lazy and reuse (if they can) the same code base to launch packages under different, unoccupied names.
}

\subsection{Dependent-hidden (RQ3)}

The key point of dependent-hidden is that attackers use a different package to camouflage or hide malicious packages to evade security detection tools.
We use dependency subgraphs to represent OSS malware: if two nodes have a dependency edge $e(u, v)$, we put them into the same subgraph.
Table~\ref{tab:dep} lists the overall dependent-hidden subgraph, including 22 groups in NPM, 13 groups in PyPI, and 3 groups in RubyGems.
We find one large subgraph in 3 OSS ecosystems (the maximum size in NPM is 232, the maximum size in PyPI is 950, and the maximum size in RubyGems is 34).
The remaining subgraphs have a small size (less than 10).
The largest subgraph is formed by multiple dependencies of packages that are reused by different malicious packages. 

\begin{table}[!t] \small
    \centering
    \caption{The details of the subgraphs based on dependency relationships.}
    \begin{tabular}{ c c c c }
    \toprule
    OSS & Pkg. Num.\#  & \makecell[c]{ Subgraph Num.\# }  & Ave. Num.\# \\
    \hline
    NPM &   323           &  22  &  14.68      \\
    PyPI &  992       &  13   & 76.31    \\
    RubyGems &  39     & 3  & 13     \\
    \toprule
    \end{tabular}
    \label{tab:dep}
\end{table}
\begin{table}[!t]
\centering
\caption{Dependency malicious packages in the subgraphs.}
\begin{tabular}{l r }
\toprule
OSS & Dependency (Num. of other malicious packages)  \\
\hline
\multirow{1}{*}{NPM} & util(88), icons(39), common(4),  object-color(3), settings(3) \\
\hline
\multirow{5}{*}{PyPI} & urllib(448), request(124), urllib3(92), timedelta(75) \\
             & values(18), public(14), pystyle(12), urlsplit(12), coloram(11)\\
            &  pwd(11), connection(10), pkgutil(10), twyne(8), runcmd(8)\\
                          
            & docutils(6), seccache(6), openvc(5), faq(4), setupcfg(4),\\
                                &  exit(4), load(3), jsfiddle(3) \\
                            
\hline
\multirow{1}{*}{Ruby} & rest-client(32) \\
    \toprule
\end{tabular}
    \label{tab:dep-num}
\end{table}

We further inspect the details of dependent-hidden subgraphs.
Table~\ref{tab:dep-num} shows the malicious packages that were reused by more than two different malicious packages.
In the NPM ecosystem, the most frequently referenced package is `util', which was claimed as a dependency library by 88 different malicious packages. In the PyPI ecosystem, `urllib' was referenced 448 times, while in the RubyGems ecosystem, `rest-client' was used as a dependency of 32 different malicious packages.
Combining Table~\ref{tab:dep} and \ref{tab:dep-num}, we observe that 1,354 malicious packages used 28 malicious packages as their dependency library.
In addition, the dependent-hidden subgraphs have two methods to hide malicious packages, as follows.
(1) Attackers tend to use common names often found in software development, such as `util(s)', `common', and `setting', as malicious package names. 
(2) Some popular legitimate packages may be embedded with malicious code in a specific version.

\begin{figure}[t!]
    \centering
    \includegraphics[width= 2.5in]{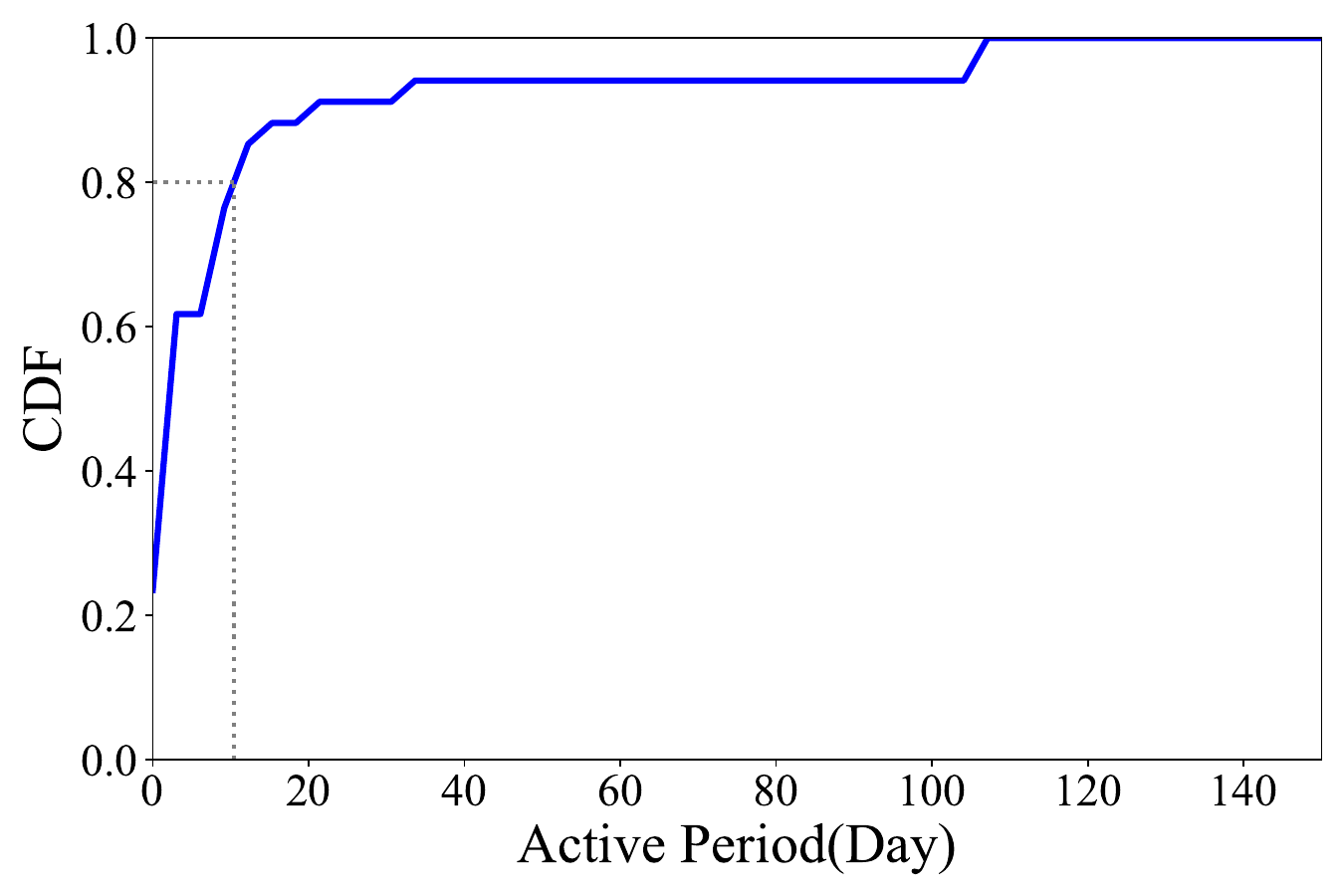}
    \caption{The active period of the subgraph based on dependency relationships.}
    \label{fig:ap:dep}
\end{figure}

\textbf{Active Time}. 
Figure~\ref{fig:ap:dep} plots the CDF of the active period of dependent-hidden subgraphs.
80\% dependent-hidden groups have less than 10 days of active periods.
It is caused by a different attack strategy.  
A dependent-hidden subgraph indicates that the attacker releases a package with malicious code and then creates a second one, depending on the first one.
Compared with similar subgraphs (45.16 days on average), the dependent-hidden subgraph has a shorter average active period (10.5 days). 
The reason is that a quick and simple attack strategy may be effective in deceiving users into downloading and installing malware.
We also find a long-tail effect of the active period for those attack campaigns, indicating some malicious packages have been released after more than 100 days.
In short, the costs of attacking OSS ecosystems by exploiting dependencies are high, while the benefits are low.

\find{
\textbf{Finding 3:} 
28 malicious packages are repeatedly hidden in attacks, some over 400 times.
In terms of active time, using dependency libraries to hide malicious packages brings little benefit compared to simple repetitive attempts.
}

\begin{table}[!t] \small
    \centering
    \caption{The details of the subgraph based on co-existing relationships.}
    \begin{tabular}{ c c c c }
    \toprule
    OSS & Pkg. Num.\#  & \makecell[c]{ Subgraph Num.\# }  & Ave. Num.\# \\
    \hline
    NPM &   3,110          &  33    &    94.24     \\
    PyPI &    7,249     &  40   &   181.23   \\
    RubyGems &   76    &  9  &  8.44     \\
    \toprule
    \end{tabular}
    \label{tab:report}
\end{table}

\subsection{Malware Context (RQ4)}

As we mentioned before, security reports disclose the information about corresponding SSC attack campaigns. 
We use the co-existing subgraphs to represent OSS malware: if two nodes have a co-existing edge $e(u, v)$, we put them into the same subgraph.
Table~\ref{tab:report} lists the overall co-existing subgraphs, including 33 subgraphs in NPM, 40 subgraphs in PyPI, and 9 subgraphs in RubyGems. 
The co-existing subgraph in PyPI contains 181.23 malicious packages on average; the average size in NPM is 94.24; and the average size in RubyGems is 8.44.
A security report may contain multiple malicious packages, indicating a many-to-one mapping between a report and packages.

\begin{figure}[!t]
    \centering
    \includegraphics[width= 2.5in]{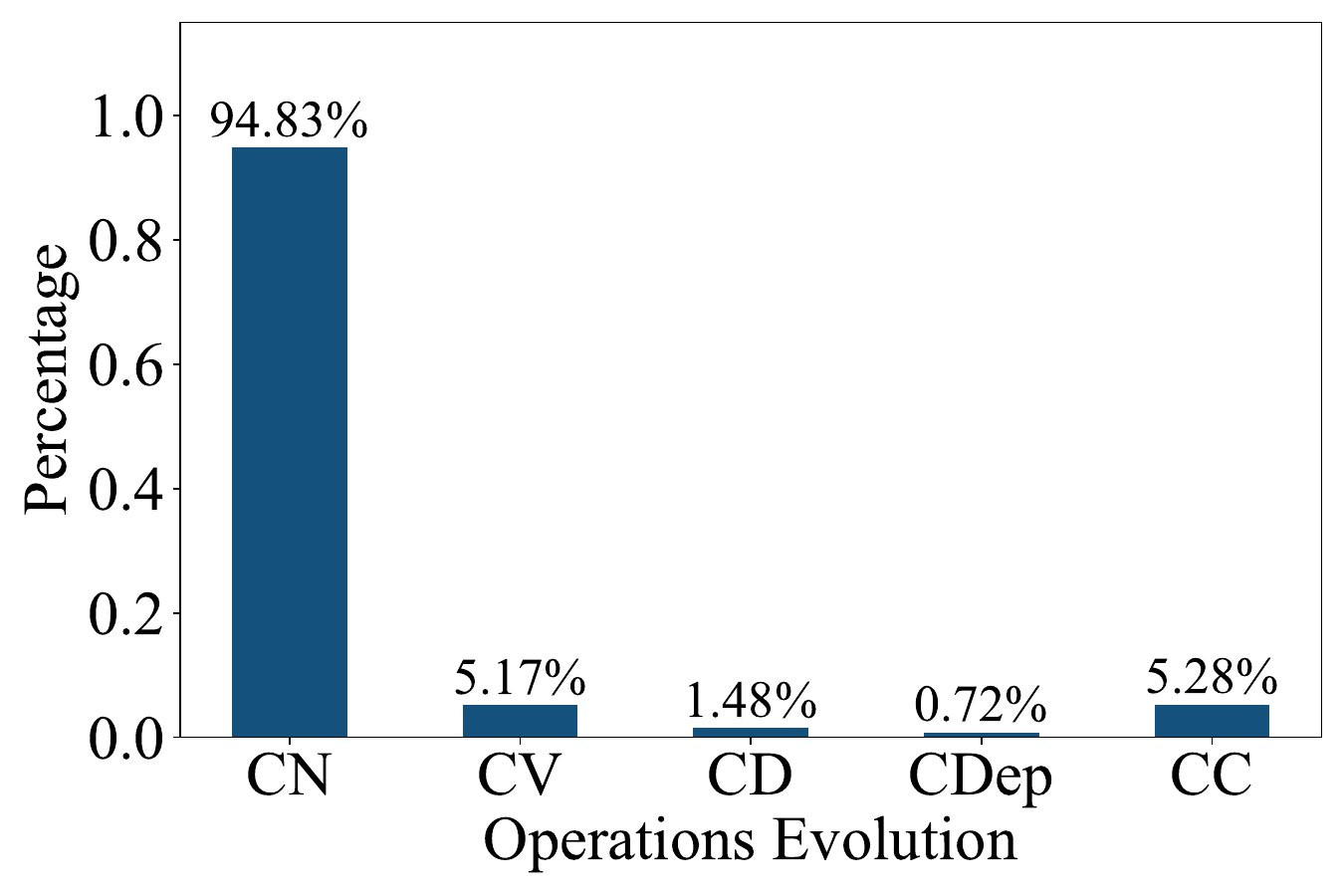}
    \caption{The operation distribution of subgraphs based on co-existing relationships.}
    \label{fig:operations evolution report}
        \vspace{-3mm}
\end{figure}

\textbf{Social Engineering Tactics}. 
Figure~\ref{fig:operations evolution report} depicts the operation distribution of subgraphs.
We observe that CN is the most popular operation, and other operations have minimal size. 
The CN's percentage is 94.83\%. 
Subsequently, the CV's percentage is 5.17\%, and the CC's percentage is 5.28\%.
It is evident that Figures~\ref{fig:operations evolution repeat} and \ref{fig:operations evolution report} are similar in the distribution of operations.
The reason is that many malicious packages from security reports belong to the malware based on similar code.

\begin{figure}[!t]
    \centering
    \includegraphics[width= 2.5in]{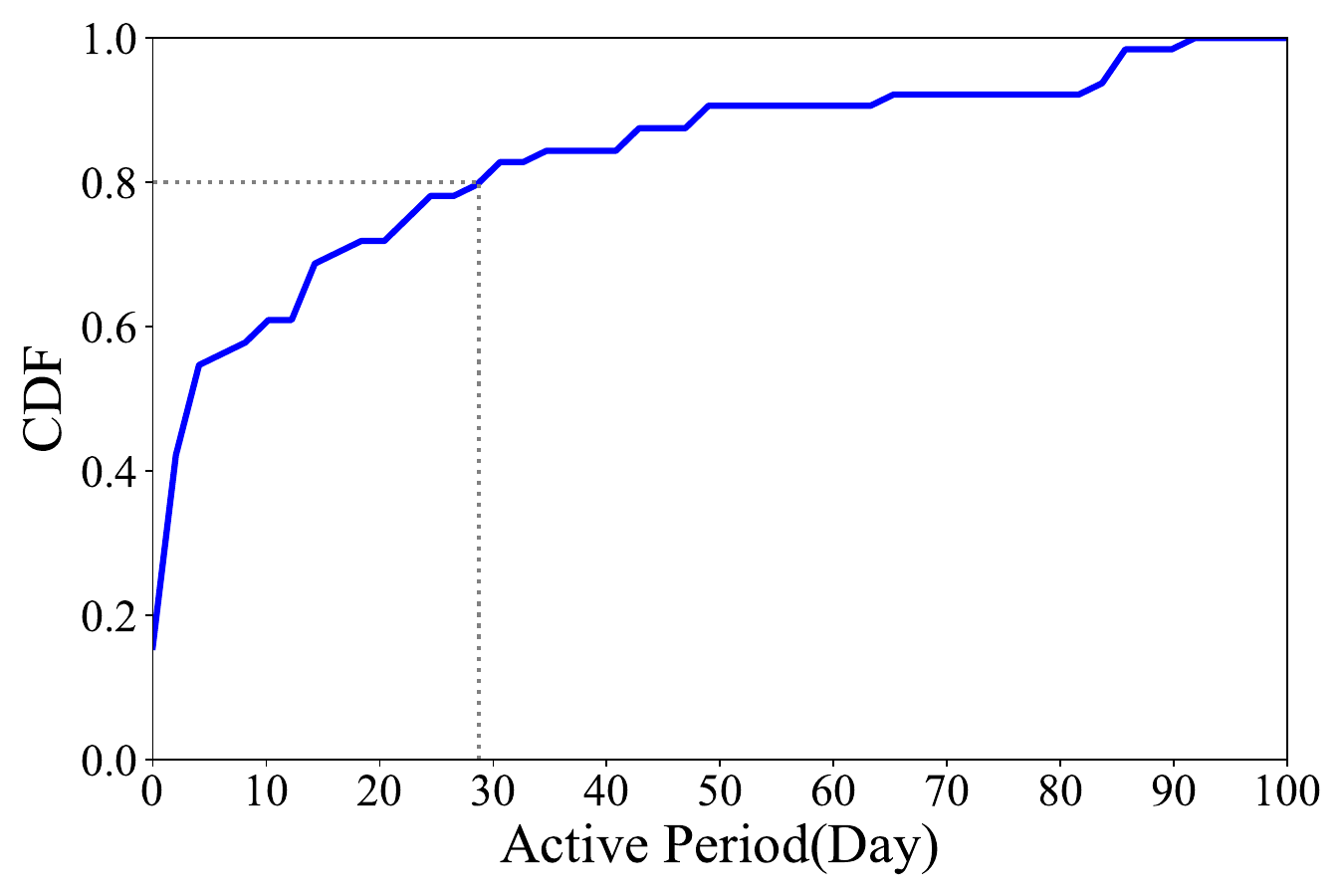}
    \caption{The active period of the subgraphs based on security reports.}
    \label{fig:ap:report}
\end{figure}

\textbf{Active Time}.
Figure~\ref{fig:ap:report} plots the CDF of the active period of subgraphs in security reports.
We observe that 20\% of the reported attacks have an active time of 0 days.
It is the reason that those malicious packages were released and taken down at the same time.
We find that the curve rises slowly, indicating that attack campaigns cover various active times. 
This is because commercial corporations and security professionals publish the newest SSC attack events as soon as possible. 
For example, Tianwen~\cite{tianwen-online} releases a security per quarter.

\textbf{Context}.
The rationale behind the reported attack is that security reports contain context about OSS malicious packages.
Security analysis reports disclose many types of attack activities. 
Analysis reports can supply relevant contextual information about attack activities, e.g., the attacker information and attack method.

\begin{figure}[!t]
    \centering
    \includegraphics[width= 2.5in]{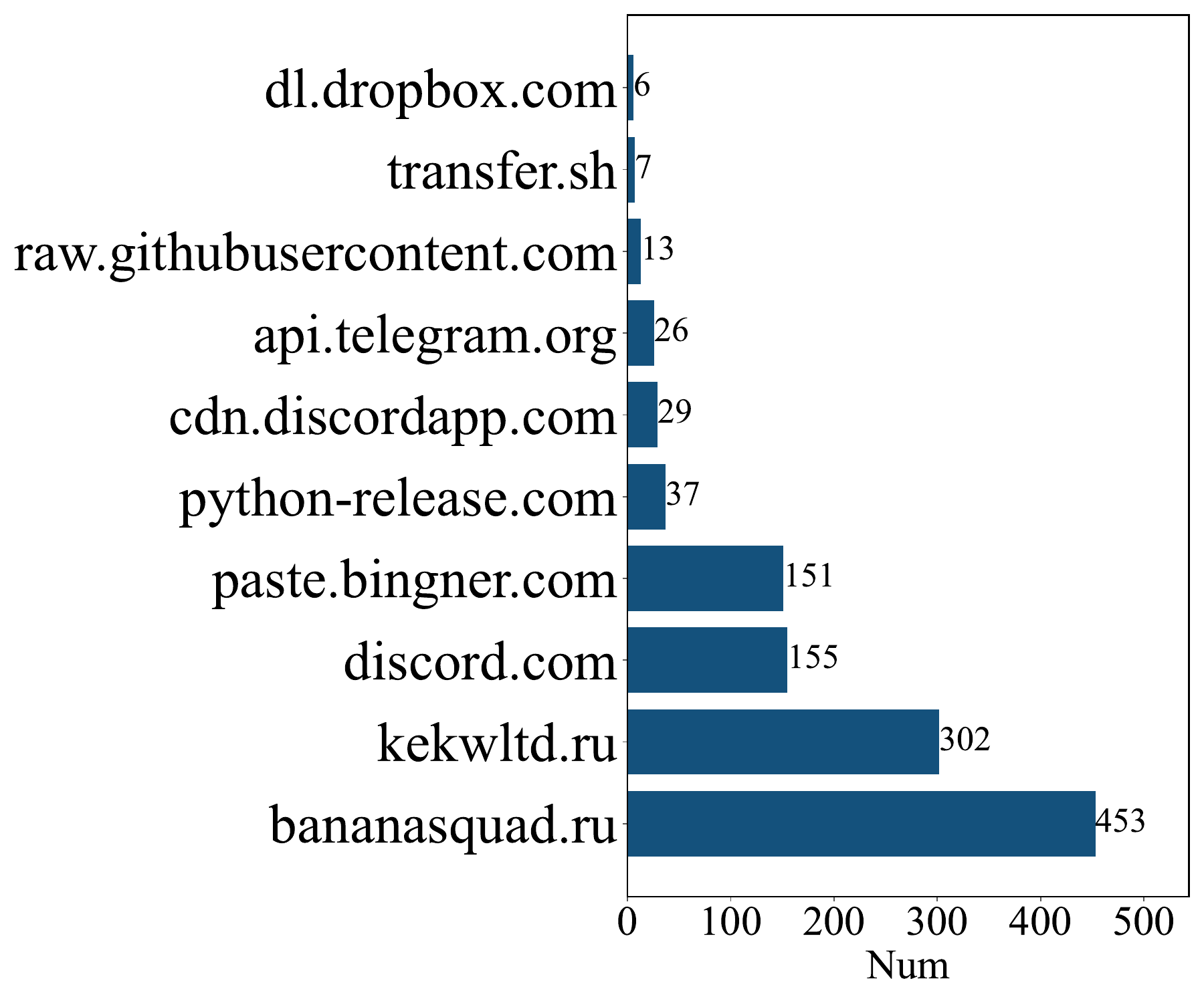}
    \caption{Top 10 of malicious domains.}
    \label{fig:url report}
        \vspace{-3mm}
\end{figure}

We extract the relevant context from the report from the subgraphs.
Specifically, we use the Indicator of Compromise (IoC) as the attack context, which is the warning sign that a system may have been breached or compromised.
Here, there are 3 types of IoC in the security reports, including suspicious IP addresses, malicious domains/URLs, and PowerShell commands/scripts.
We counted the details of IoC indicators from our reported attack campaigns. 
There are 1,449 malicious URLs, 234 suspicious IP addresses, and 4 PowerShell commands. 
Note that those IoCs belong to the forensic data that security professionals use to detect potential malicious activity on a network or system.

We preprocess these IoCs, including removing the duplicate ones and identifying available ones.  
We explored IP addresses among 234.
The maximum number of the same IP appears is 23 (46.226.*.*, 51.178.*.*, 81.24.*.*, 141.95.*.*, 135.181.*.*, 195.201.*.*, 5.135.*.*). 
We manually identify whether those IP addresses are available, and the results are that all IP addresses cannot be accessible.
Figure~\ref{fig:url report} shows the top 10 domains among 1,449 URLs.
The domain `bananasquad.ru' has the largest number, 453 times, followed by `kekwltd.ru' with 302 times.
Those domains may belong to Command and Control Infrastructure, as C2, which is the server that attackers use to communicate with malicious packages for following initial exploitation. 
Malware uses them for remote control, data theft, or updating malicious instructions.
We find that some URLs are used by malware to download executable files (such as .exe or .dll), or a malicious payload such as a script file or tool. 

\find{
\textbf{Finding 4:} 
Many malicious packages lack the context of the attack campaign, and the security report is the evidence to reveal the attack campaign behind the malware.
}

\section{Security Implication and Application}

This section demonstrates the security implications and applications of {\tech}.

\subsection{Malware Detection}

\begin{table}[!t] \small
\caption{{\color{black}The performance of malware detection with and without {\tech}. ``w/o'': without, ``w'': with.}}
\centering
\begin{tabular}{c|cc|cc}
\toprule

Algorithm & \makecell{Acc.\\(w/o)} & \makecell{Acc.\\(w)} & \makecell{Recall\\(w/o)} & \makecell{Recall\\(w)} \\
\midrule
{\color{black}RF}   & 0.897 & 0.944\textsuperscript{+5.2\%}  & 0.825 & 0.984\textsuperscript{+19.3\%} \\
{\color{black}LR}   & 0.841 & 0.859\textsuperscript{+2.1\%}  & 0.806 & 0.836\textsuperscript{+3.7\%}  \\
{\color{black}KNN}  & 0.773 & 0.807\textsuperscript{+4.4\%}  & 0.778 & 0.818\textsuperscript{+5.1\%}  \\
{\color{black}MLP}  & 0.860 & 0.895\textsuperscript{+4.1\%}  & 0.839 & 0.927\textsuperscript{+10.5\%} \\
\bottomrule
\end{tabular}
\vspace{-2mm}
\label{tab:accuracy_comparison}
\end{table}

{\color{black}
As we mentioned before, many malicious packages are centered on a few aggressive packages that dominate the dataset.
To build a classification model based on our dataset, we apply both machine learning and deep learning algorithms.  
In one category, we use the malware diversity to train the model; but in the other category, we do not use the malware diversity to train the model. 
Specifically, we evaluate the performance of three machine learning algorithms (Random Forest (RF) and Logistic Regression (LR),  and K-Nearest Neighbors (KNN)) alongside one simple deep neural network (Multi-Layer Perceptron (MLP)).

}

The dataset includes all 3,840 tracked malware packages in NPM and 3,500 random legitimate packages from the registry.
In each iteration, we randomly select a fixed number of clusters and sample two packages from each cluster into the test set, allowing repeated sampling for small clusters.
The remaining samples are used to construct training sets in two ways:
(1) we sample two packages from each cluster;
(2) we randomly sample the same number of malicious packages from the remaining pool without using any clustering information.
In both settings, we sample an equal number of legitimate packages for balance.
We repeat this process for 50 iterations.
We implement all classifiers using scikit-learn~\cite{scikit} and Pytorch~\cite{pytorch}.
Table~\ref{tab:accuracy_comparison} lists the average performance over 50 iterations of malware detection.
With the help of malware diversity, the model achieves improvements in overall performance, with an average recall increase of around 10\%.
In short, {\tech} can improve the performance of malware detection.

\subsection{Malware Analysis}

\begin{table}[!t]
\small
\centering
\caption{{\color{black}The quantitative analysis of malicious behaviors.}}
\label{tab:cluster_behavior}
\begin{tabular}{l c c}
\toprule
         &  {\color{black}Pkg. Num.} &  {\color{black}Malicious Behavior}  \\
\midrule
\multirow{14}{*}{NPM} & 827 &  {\color{black}\makecell[l]{Spyware, Backdoor, Data exfiltration \\ via TLS}} \\
 & 414 &  {\color{black}\makecell[l]{C2 channel,  Credential collecting \\ DNS tunneling exfiltration} }\\
 & 196 &  {\color{black}\makecell[l]{Beaconing, Fingerprint Spoofing \\C2 channel}}\\
 & 149 &  {\color{black}\makecell[l]{Webhook Abuse, Surveillance \\ Data Exfiltration} }\\
 & 118 &  {\color{black}\makecell[l]{Webhook Abuse, Fingerprinting\\ Service Abuse}} \\
 & 118 &  {\color{black}\makecell[l]{Beaconing, UA Spoofing,  C2 channel}} \\
 & 118 &  {\color{black}\makecell[l]{Identity Exfiltration, Data Exfiltration}} \\ 
 & 110 &  {\color{black}\makecell[l]{Data Exfiltration, PII collecting \\ OAuth2 Abuse}} \\
\midrule
 
\multirow{10}{*}{PyPI} & 829 &  {\color{black} \makecell[l]{Chinese Character Obfuscation \\ Network Behavior Masking} }\\

 & 409 &   {\color{black}\makecell[l]{Discord Payload Delivery, PowerShell \\Legitimate Package Spoofing} }\\
 & 270 &   {\color{black}\makecell[l]{Dropbox Malware Fetch,  License Abuse} }\\
 & 180 &   {\color{black}\makecell[l]{Obfuscation, Legitimate Package Spoofing}} \\
& 140 &  {\color{black} \makecell[l]{PowerShell, Legitimate Package Spoofing} }\\
 & 134 &  {\color{black} \makecell[l]{Dropbox Malware Fetch, PowerShell \\Legitimate Package Spoofing}} \\

\bottomrule
\end{tabular}
\end{table}

{\tech} can help security professionals to obtain a deep understanding of malicious behaviors. 
We provide a quantitative analysis of the malicious behaviors of OSS malware.
Specifically, we leverage the similar edge in {\tech} to obtain OSS malware groups.
If the group size is larger than 100, we inspect malware code to characterize its behaviors. 
The analysis approach is as follows. 
(1) If the malware is reported by online sources (Table~\ref{tab:report:source}), we use the security report content to represent its behaviors.
(2) If the malware is not reported, we use GPT-4o~\cite{openai} to describe its behaviors. Then, we further inspect its code to avoid LLM's hallucinations and errors.
(3) All malicious behavior labels are curated by individual experience. 

{\color{black}
Table~\ref{tab:cluster_behavior} lists the quantitative analysis of malicious behaviors across NMP and PyPI. 
The results show that malicious packages in NPM primarily exhibit two behaviors: data exfiltration and obfuscation. 
Similarly, malicious PyPI packages disguise themselves as legitimate tools by using spoofed MIT licenses and development classifiers, evading detection with multi-layered obfuscation like Base64-encoded commands and hidden PowerShell windows.
For example,  attackers use unique characters and their combinations to achieve the obfuscation~\cite{malware-chinese-code}.
}

\section{Discussion \& Limitation}

\textbf{Malware Diversity} plays a vital role in conventional malware research~\cite{rossow2012prudent}, guaranteeing the validity of the research.
Each conventional malware has a lot of samples in binary form, and each sample is different from others at the binary level, called binary polymorphism.
However, each OSS malware is released as a software package in the OSS ecosystem (without binary-level samples), including its metadata, dependencies, versions, and source code.
The diversity of OSS malware is different from that of conventional malware, and we cannot use the malware family to represent it.  
So far, the diversity of OSS malware is unclear. 
It will be a future work to provide a new definition of the OSS malware diversity.

\textbf{Missing Malware Packages}.
As we mentioned before, many malware packages are missing from our dataset. 
Once the malware is removed, there is a low probability that we can find it at other online sources.
The only way to make amends for those malware packages is to cooperate between the industry and academia. 
We will release our malware corpus to the public, and we hope the community can help us supplement the missing packages.

\textbf{Analysis Results Instability}.
One concern is that our analysis results may change over time when new malware packages are released or missing packages are supplemented.
Indeed, the increasing number of malware packages may affect the result of the analysis.
In our dataset, the release timeline of the malicious packages covers from 2014 to 2024.
Hence, our analysis results are stable with time.

{\color{black}
\textbf{Data Representativeness}.
One concern is whether our dataset can reflect the whole OSS malware landscape.
Due to distinct distribution channels, currently there lack of such a representative dataset. 
So far, our dataset has several advantages in acting as a representative dataset. 
(1) Our dataset involves the largest number of OSS malware compared with other existing datasets~\cite{duan2020towards, guo2023empirical, Ohmssc2020}. 
(2) We gather malware packages from a variety of online sources, ensuring a broad representation of the malware landscape.
(3) We have cross-referenced our dataset with several prior studies to validate its relevance and comprehensiveness.
(4) As shown in Figure~\ref{fig:untrack:timeline}, our dataset spans multiple years, encompassing both emerging and established threats, which strengthens its representativeness over time.
}

\section{Related Work}

\textbf{Malware Analysis} is to obtain a deep understanding of malware behaviors. 
Prior work on binary malware~\cite{bailey2007automated, bayer2009scalable, jang2011bitshred} leveraged the malware samples with traces to picture the malware behaviors. 
Bayer et al.\cite{bayer2009scalable} and Jiang et al.\cite{jang2011bitshred} proposed clustering techniques to partition malware samples into different groups based on their behaviors, and they analyzed the behavioral profiles of malware samples.
VX underground~\cite{vx_underground} is the largest open database for binary malware, 
which contains a large collection of malware samples and corresponding code bases. 
Rokon et al.\cite{rokon2020sourcefinder} proposed SourceFinder to search the GitHub repository based on malware samples and provided a dataset of malware codes.
Those databases are designed for binary malware, and they are not suitable for OSS malicious packages.

In contrast, the research community has turned to OSS malware analysis, including the malware detection~\cite{cappos2008look, SejiaAdria2022Machinelearn, zhang2020cyber, qian2022malicious, ferreira2021containing, vu2023bad}, the malware behaviors~\cite{wy2022InstalltimeAtt, lidisa2022javabytecode, sejfia2022practical}, and the malware dataset~\cite{backstabbers-online, Ohmssc2020, guo2023empirical}.
Pfretzschner et al.~\cite{pfretzschner2017identification} proposed a detection algorithm for dependency-based attacks on Node.js, and Staicu et al.~\cite{staicu2018understanding} proposed a deep understanding of the injection attack on Node.js.
Cao et al.~\cite{cao2022fork} leveraged the artifacts to find 26 fork repositories among 68 popular cryptocurrency repositories.
Several OSS malware datasets are presented in \cite{duan2020towards, guo2023empirical, Ohmssc2020}, whose package number is smaller than ours.

\textbf{OSS Ecosystem} is a complex network of software packages and developers. 
Decan et al.~\cite{decan2017empirical} and Zimmermann et al.~\cite{zimmermann2019small} study that the package dependency has become more complicated as time passed, and the package security risk is also increasing, which can serve as a launching pad for attackers.  
Dann et al.~\cite{dann2021identifying} identified dozens or hundreds of the direct and transitive dependencies.
Vasilakis et al.~\cite{vasilakis2021supply} proposed a vetting framework for reused components in internal/public repositories.
Typosquatting is the most popular attack vector in the OSS ecosystem~\cite{vu2021lastpymile, taylor2020spellbound}. 
Xiao et al.~\cite{xiao2021abusing} proposed an attack that abuses hidden attributes, which attackers can exploit to obtain confidential data, bypass security checks, and launch denial-of-service attacks. 
Pashchenko et al.~\cite{pashchenko2020qualitative} claimed that many developers are unaware of dependency issues or do not want to modify the software. 
Zahan et al.~\cite{zahan2021weak} analyzed the metadata of 1.63 million packages and provided six indicators of security risks in the NPM ecosystem.
Li et al.~\cite{li2021dpg} proposed the dependency graphs between packages for analyzing known vulnerability impacts or underlying risks among different dependencies. 
Gu et al.~\cite{gu2023investigating} built RScouter and found 12 potential attack vectors in the 6 registries, which can be used to distribute malicious packages. 
Wyss et al.~\cite{wyss2022wolf} proposed three SSC attacks, including install-time, import-time, and run-time attacks.

\section{Conclusion}

In this paper, we build the largest OSS malware dataset from scattered online sources, including 24,000+ malicious packages.
Further, we propose a knowledge graph to represent the OSS malware corpus, which security professionals can use to develop and evaluate malware detection and analysis approaches.
Our quantitative analysis 
provides four insights into the research community: (1) OSS malware analysis research still needs to collect and clean the dataset;
(2) the OSS malware diversity is an unsolved problem, causing noises and errors; (3) only 28 malicious packages were repeatedly hidden via dependency libraries; (4) the context of malware-based attacks are essential artifacts for today's security community.

\section*{Acknowledgement}

We are grateful to our shepherd Hiroshi Yamada and anonymous reviewers for their insightful feedback. 
The work was partially supported by the National Natural Science Foundation of China (No. 62272029 and No. 61972024).

\small
\bibliographystyle{IEEEtran}
\bibliography{ref/ref_software_supply_chain, ref/ref_online, ref/ref_algorithm, ref/ref_llm, ref/ref_ours}

\end{sloppypar}
\end{document}